\documentclass[nocopyrightspace]{sigplanconf}

\usepackage{multirow}
\usepackage{etex}
\usepackage{marginnote}
\usepackage{amsmath, amssymb, amsfonts}
\usepackage{amsthm}
\usepackage{paralist}
\usepackage{alltt}
\usepackage{wrapfig}
\usepackage{stmaryrd}
\usepackage{MnSymbol}
\usepackage{paralist}
\usepackage{booktabs}

\usepackage{thmtools,thm-restate}
\usepackage[usenames,dvipsnames]{xcolor}
\usepackage{caption}
\usepackage{subcaption}
\usepackage{floatrow}
\usepackage{calc}% To calculate width for \FBwidth
\usepackage{algorithm}
\usepackage[noend]{algpseudocode}

\newtheorem{theorem}{Theorem}[section]

\usepackage{tikz}
\usetikzlibrary{matrix}
\usetikzlibrary{fit}
\usetikzlibrary{automata}
\usetikzlibrary{shapes,backgrounds}

\newcommand\comment[1]{}

\usepackage{xspace}
\usepackage{array}
\usepackage{proof}

\newcommand\arsays[1]{}

\newcommand\ttsays[1]{}
\newcommand\lrsays[1]{}
\newcommand{\rs}[1]{}

\newcolumntype{t}{>{\tt}l}

\newcommand{\hbra}{
  \hbox to \columnwidth{\vrule width0.3mm height 1.8mm depth-0.3mm
    \leaders\hrule height1.8mm depth-1.5mm\hfill
    \vrule width0.3mm height 1.8mm depth-0.3mm}}
\newcommand{\hket}{
  \hbox to \columnwidth{\vrule width0.3mm height1.5mm
    \leaders\hrule height0.3mm\hfill
    \vrule width0.3mm height1.5mm}}

\makeatletter
  \newcommand{\addToLabel}[1]{%
    \protected@edef\@currentlabel{\@currentlabel#1}%
  }
\makeatother
\newcommand{\ratio}{.35}

\newcounter{excounter}

\newenvironment{display}[2][\ratio]{\vspace{-0.5ex}
\begin{tabbing}
     \=  \=  \= \kill
    \textbf{#2}\\[-.8ex]
    \hbra\\[-.8ex]
  }{\\[-.8ex]\hket
  \end{tabbing}\vspace{-1ex}}

\newcounter{rule}

\newcommand{\staterule}[4][]{%
  \refstepcounter{rule}%
  \addToLabel{(#2)}%\label{#2}%
  $\begin{array}[b]{@{}l@{}}%
    \begin{array}{@{}c@{}}
      #3\\
      \hline
      \raisebox{0ex}[2.5ex]{\strut}#4%
    \end{array}
    \mbox{{\sc #2} #1}\\%
  \end{array}$}

\newcommand{\GAP}{1ex}

\newcommand{\out}{\mathcal{O}}

\newcommand{\smallstepx}[3]{\langle #1 \rangle \xrightarrow{#3} \langle #2 \rangle}

\newcommand{\smallstep}[2]{\smallstepx{#1}{#2}{\epsilon}}

\newcommand{\kw}[1]{\ensuremath{\mathtt{#1}}}

\newcommand{\e}{e}

\newcommand{\Asgn}[2]{#1:=#2}

\newcommand{\Input}[1]{\kw{in}(#1)}
\newcommand{\Output}[2]{\kw{out}(#1,#2)}

\newcommand{\Havoc}{{\tt havoc}}

\newcommand{\VariableValuation}{\calV}
\newcommand{\evalExpression}[1]{#1[v / \VariableValuation[v]]}

\newcommand{\secref}[1]{Sec.~\ref{sec:#1}}

\newcommand{\figref}[1]{Fig.~\ref{fig:#1}}

\newcommand{\src}[1]{\texttt{#1}}

\newcommand\calC{\mathcal{C}}

\newcommand\calV{\mathcal{V}}

\newcommand\instr{stmt}
\newcommand\tid{\mathit{tid}}

\newcommand\techReport{false} %TODO use this command for normal paper
\newcommand\ifTechReport[1]{\ifthenelse{\equal{\techReport}{true}}{#1}{}}
\newcommand\ifPaper[1]{\ifthenelse{\equal{\techReport}{true}}{}{#1}}

\newcommand{\Lang}[1]{\ensuremath{\mathcal L(#1)}}

\newcommand\ProgState{\ensuremath{s}}

\newcommand\natset{\ensuremath{\mathbb{N}}}

\newcommand{\loc}{\ell}
\newcommand{\wlang}{\mathcal W}
\newcommand{\wlangabs}{{\mathcal W}_{abs}}

\newcommand{\nfa}{NFA}
\newcommand{\nfas}{NFAs}

\newcommand{\upto}{modulo}
\newcommand{\suc}{\mathtt{succ}}

\newcommand{\ourtool}{{\sc O-Liss}}
\newcommand{\pasttool}{{\sc Liss}}

\newcommand{\hbformula}{HB-formula}

\newcommand{\Naturals}{\mathbb{N}}

\newcommand{\thread}{\mathtt{T}}
\newcommand{\cProg}{\calC}

\newcommand{\sem}[1]{[\![ #1 ]\!]}

\newcommand{\mtt}[1]{\mathtt{#1}}

\newcommand\cfg{\cal G}

\newcommand\Reals{\mathbb{R}}
\newcommand\wprog{\mathbb{W}}
\newcommand\wprogabs{\mathbb{W}_{abs}}

\newcommand{\guard}{\texttt{g\_var}}
\newcommand{\last}{{\tt last}}

\newcommand{\stmt}[1]{\mathtt{stmt}({#1})}

\newcommand{\cProgabs}{\cProg_{abs}}
\newcommand{\autNP}{\mathsf{NP}_{abs}}
\newcommand{\autP}{\mathsf{P}'_{abs}}
\newcommand{\Conflicts}{\mathbb{C}}
\newcommand{\AllConflicts}{\Conflicts_{\text{all}}}

\newcommand{\new}{\text{new}}
\newcommand{\old}{\text{old}}
\newcommand{\pre}{\text{pre}}
\newcommand{\post}{\text{post}}

\newcommand{\conpre}{\text{cpre}}
\newcommand{\conpost}{\text{cpost}}
\newcommand{\mi}{\text{mid}}
\newcommand{\n}{K}
\newcommand{\x}{x}
\newcommand{\lk}{lk}

\newcommand{\smt}{\mathsf{LkCons}}

\newcommand{\pred}{\mathsf{Pre}}
\newcommand{\Locks}{\mathsf{Lk}}
\newcommand{\Loc}{\mathsf{L}}
\newcommand{\ord}{\mathsf{Order}}

\newcommand{\freq}{\mathtt{freq}}
\newcommand{\cost}{\mathtt{cost}}

\newcommand{\LockCost}{tl}

\newcommand{\block}{\mathtt{block}}

\newcommand{\InLock}{\mathsf{InLo}}
\newcommand{\LockBefore}{\mathsf{LoBef}}
\newcommand{\LockAfter}{\mathsf{LoAft}}
\newcommand{\UnlockBefore}{\mathsf{UnBef}}
\newcommand{\UnlockAfter}{\mathsf{UnAft}}
\newcommand{\InLockExt}{\mathsf{InLoEnd}}

\newcommand{\AbsMin}{\mathsf{Minimal}}
\newcommand{\Coarsest}{\mathsf{Coarse}}
\newcommand{\Finest}{\mathsf{Fine}}
\newcommand{\PerfModel}{\mathsf{PerfModel}}

\newcommand{\InstrSet}{S}

\newcommand{\ITE}[3]{\mathtt{if}(#1)~\mathtt{then}~#2~\mathtt{else}~#3}

\newcommand{\Region}{\mathit{Reg}}
\newcommand{\Regions}{\mathit{Regs}}

\newcommand{\UniquePerformance}{\mathit{UniquePerformance}}
\newcommand{\MinPerf}{\mathit{Min}}
\newcommand{\MinModel}{\mathit{Model}}

\newcommand{\blk}{b}
\newcommand{\criticalCost}{tc}
\newcommand{\contention}{\kappa}
\newcommand{\LocksAcquired}{\nu}

\usepackage[plainpages = false, pdfpagelabels, pdfborder={0 0 0}, pdfdisplaydoctitle=true,pdflang=en-gb,pdfencoding=auto,bookmarksnumbered=true,bookmarks=true,bookmarksopen=true,pdfstartview={XYZ null null 1}]{hyperref}

\conferenceinfo{CONF 'yy}{Month d--d, 20yy, City, ST, Country} 
\copyrightyear{20yy} 
\copyrightdata{978-1-nnnn-nnnn-n/yy/mm} 
\doi{nnnnnnn.nnnnnnn}

\begin{document}
\title{Optimizing Solution Quality in Synchronization Synthesis
%\vspace{-4ex}
}
%\authorinfo{}{}{}
 \authorinfo{Pavol {\v C}ern{\'y}}{University of Colorado Boulder}{pavol.cerny@colorado.edu}
 \authorinfo{Edmund M. Clarke}{Carnegie Mellon University}{emc@cs.cmu.edu}
 \authorinfo{Thomas A. Henzinger}{IST Austria}{tah@ist.ac.at}
 \authorinfo{Arjun Radhakrishna}{University of Pennsylvania}{arjunrad@seas.upenn.edu}
 \authorinfo{Leonid Ryzhyk}{Carnegie Mellon University}{ryzhyk@cs.cmu.edu}
 \authorinfo{Roopsha Samanta}{IST Austria}{rsamanta@ist.ac.at}
 \authorinfo{Thorsten Tarrach}{IST Austria}{ttarrach@ist.ac.at}

\maketitle

\begin{abstract}
%\rs{Replace globally: synchronization with locks as appropriate}
Given a multithreaded program written assuming a friendly, non-preemptive scheduler,                     
the goal of synchronization synthesis is to automatically
insert synchronization primitives to ensure that the modified program behaves 
correctly, even with a preemptive scheduler. 
In this work, we focus on the quality of the synthesized solution: we aim to 
infer synchronization placements that not only ensure correctness, but 
also meet some quantitative objectives such as optimal program performance 
on a given computing platform.

The key step that enables solution optimization is the construction of 
a set of global constraints over synchronization placements  
such that each model of the constraints set corresponds to a
correctness-ensuring synchronization placement.  
We extract the global constraints from generalizations of 
counterexample traces and the control-flow graph of the program.
The global constraints enable us to choose from among the  
encoded synchronization solutions using an objective function. 
We consider two types of objective functions: ones that 
are solely dependent on the program (e.g., minimizing the size of
critical sections)  
and ones that are also dependent on the computing platform. 
For the latter, given a program and a computing platform, 
we construct a performance model based on measuring average contention 
for critical sections and the average time taken to acquire and release a
lock under a given average contention. 

We empirically evaluated that our approach scales to
typical module sizes of many real world concurrent programs such as
device drivers and multithreaded servers, and that the 
performance predictions match reality. 
To the best of our knowledge, this is the first comprehensive approach 
for optimizing the placement of synthesized synchronization. 
\end{abstract}

\section{Introduction}
\label{sec:intro}

Synchronization synthesis aims to enable programmers to concentrate on
the functionality of the program, and not on the low-level
synchronization. One of the main challenges in synchronization
synthesis, and in program synthesis in general, is to produce a
solution that not only satisfies the specification, but also is good
(if not optimal) according to various metrics such as performance and
conformance to good programming practices. Optimizing for performance
for a given architecture is a challenging problem, 
but it is an area where 
program synthesis can have an advantage over the traditional approach to
program development: the fact that synthesis takes a high-level
specification as input gives the synthesizer a lot of freedom to find
a solution that is both correct and performs well.  

Optimization is hard for the common approach to synchronization
synthesis, which implements a counterexample-guided inductive
synthesis (CEGIS) loop. A counterexample is given by a trace, and
typically, this trace is immediately (greedily) removed from the
program by placing synchronization. The advantage of this trace-based
approach is that traces are computationally easier to analyze. 
However, the greedy approach makes it difficult to optimize the
final solution with respect to qualities such as performance.  

We propose a new approach that keeps the trace-based technique, but uses
it to collect a set of {\em global constraints} over synchronization
placements. Each model of the global constraints corresponds to a
correct synchronization placement.  Constructing the global constraints
is a key step in our approach, as they enable us to choose from
among the encoded synchronization solutions using an {\em objective
  function}. 

The global constraints are obtained by analyzing the program with
respect to the concurrency specification. In this paper, we focus
solely on one type of synchronization -- locks.
Our concurrency specification consists of three parts: preemption-safety, 
deadlock-freedom, and a set of standard locking discipline conditions. 
The goal of preemption-safety (proposed
in~\cite{CAV15}) is to enable the programmer to program assuming a
friendly, non-preemptive scheduler. It 
is then the task of the synthesizer to ensure that every
execution under the preemptive scheduler is observationally
equivalent to an execution under the preemptive scheduler. 
Two program executions are observationally equivalent if they generate
the same sequences of calls to interfaces of interest. 
We consider a program correct if, in addition to preemption-safety, it 
does not produce deadlocks, and if it follows good programming
practices with respect to locks: we require no double locking, no
double unlocking and several other conditions that we refer to as
legitimate locking. Legitimate locking helps making the
final solution readable and maintainable. 
The salient point of our correctness notion is that it is generic ---
the programmer does not need to write a specification for each
application separately. 

The global constraints resulting from the preemption-safety
requirement are obtained from an analysis of generalized counterexamples. 
The analysis uses the CEGIS approach
of~\cite{CAV15}, where the key steps in checking preemption-safety are
a coarse, data-oblivious abstraction (shown to work well for systems
code), and an algorithm for bounded language inclusion checking. 
The approach
of~\cite{CAV15} is greedy, that is, it immediately places
locks to eliminate the counterexample.
In contrast, we do not place locks, but instead infer 
{\em mutual exclusion (mutex) constraints} for eliminating the 
counterexample. We then enforce these
mutex constraints in the language inclusion check to avoid getting the
same counterexample again. We accumulate the mutex constraints from all  
counterexamples iteratively generated by the language inclusion
check. Once the language inclusion check succeeds, we construct 
the set of global constraints using the accumulated mutex constraints 
and constraints for enforcing deadlock-freedom and legitimate locking.

%% - Gotos: why they are there and why they cannot be removed (do not want to change program significantly). 
%% And how this makes the lock placement harder.

%% =====
Given the global constraints, we can choose
from among the encoded solutions using an objective function. 
We consider two types of objective functions: ones that 
are solely dependent on the program   
and ones that are also dependent on the computing platform and workload. 
%For the latter, given a program and a computing platform, 
%we construct a performance model based on measuring average contention 
%for critical sections and the average time taken to acquire and release a
%lock under a given average contention. 
%
%
%
%We
%consider two objective functions. First, objective functions that are
%architecture independent, that is, ones that are solely dependent on the
%semantics of the program. Second, we consider objective functions that
%depend on the target machine architecture, such as performance.  
Examples of objective functions of the first type include 
minimizing the number of {\tt lock} statements (leading to
coarse-grained locking) and maximizing concurrency 
(leading to fine-grained locking). We encode such an objective function, together
with the global constraints, into a weighted MaxSAT
problem, which is then solved using an off-the-shelf solver. 

In order to choose a lock placement (from among the ones
encoded by global constraints) that has a good performance, we use an
objective function that depends on a particular machine
architecture and workload. The objective function is given by a performance model.
We emphasize that the model is based on measuring parameters of a
particular architecture running the program. In particular, it is  
based on measuring the average time taken to acquire and release a
lock under a given level of contention (this part depends
only on the architecture) and the average time it takes to execute the
critical section, and the average contention 
for critical sections (this part depends on the architecture, the
program, and the workload). The optimization procedure using the performance model and
the global constraints works as follows. 
%First, we obtain the
%maxmimally coarse-grained solution of the global constraint C and a
%finer-grained solution D (we do not take the most fine-grained
%solution as it leads to small blocks of code which are difficult to
%measure correctly). This is semi-automated, even though in principle
%could be automated well by choosing a machine independent objective
%function. 
First, we parameterize the
space of solutions by the sizes of critical sections and the number of
locks taken. Second, we find the parameter values that maximize
performance. Third, we find a solution of the global constraints
closest to the parameter values that yield maximal performance. 

We empirically evaluate that our approach scales to
typical module sizes of many real world concurrent programs such as
device drivers and multithreaded servers ($\sim$1000 LOC).  
%We evaluated the performance model.
We use our synthesis 
tool (with architecture-independent objective functions) 
on a number of device driver benchmarks and find that the synthesis
times are comparable to an existing tool~\cite{CAV15} that
implements a standard CEGIS-based algorithm; we emphasize that our tool 
finds an optimal lock placement that guarantees preemption-safety, 
deadlock freedom and legitimate locking.
%such as minimal size of critical sections. 
%
Furthermore, we evaluate the tool with an objective function given by
our performance model. We use the memcached network server that
provides an in-memory key-value store, specifically the module used by 
server worker threads to access the store. We found that the performance model
predictions match reality, and that we obtained different locking
schemes based on the values of parameters of the performance model. 

\begin{figure}[t]
\begin{alltt}
static void* worker\_thread(void *arg) \{
    for (j = 0; j < niter; j++) \{
        sharedX();
        sharedY();
        local();
        sharedZ();
\}  \};
\end{alltt}
\vspace*{-0.6cm}
\caption{Example: Work sharing}
\label{ex:worksharing}
\end{figure}

The contributions of this work are:
\begin{compactitem}
\item To the best of our knowledge, this is the first comprehensive approach 
for finding and optimizing lock placement. The
approach is comprehensive, as it fully solves the lock 
placement problem on realistic (albeit simplified) systems code. 
\item We use trace analysis to obtain global constraints each of whose 
  solutions gives a legitimate lock placement, thus enabling
  choosing from among the encoded correct solutions using an objective
  function.  
\item A method for lock placement for machine-independent objective
  functions, based on weighted MaxSAT solving. 
\item Optimization of lock placement using a performance
  model obtained by measurement and profiling on a given platform. 
\end{compactitem}

\subsection{Related Work}

Synthesis of lock placement is an active research
area~\cite{bloem,VYY10,CCG08,EFJM07,UBES10,DR98,VTD06,ramalingam,SLJB08,ZSZSG08,shanlu,CAV11,CAV13,CAV14,CAV15,POPL15}.     
There are works that optimize lock placement:
for instance the paper by Emmi et al.~\cite{EFJM07}, Cherem et
al.~\cite{CCG08}, and Zhang et
al.~\cite{ZSZSG08}. These papers takes as an input
a program annotated with atomic sections, and 
replace them with locks, using several types of fixed objective
functions. 
In contrast, our work does not need the annotations with atomic
sections, and does not optimize using a fixed objective function,
but rather using a performance model obtained by performing measurements on a
particular architecture.

Another approach is to not require the programmer to annotate the
program with atomic sections, but rather infer them or infer locks
directly~\cite{VYY10,bloem,DR98,CAV13,CAV14,CAV15,POPL15}. All these
works also either do not optimize the lock placement, or
do so for a fixed objective function, not for a given
architecture. The work~\cite{CAV11} proposes concurrency synthesis
w.r.t. a performance model, but it does not produce such models for a
given machine architecture.

Jin et a~\cite{shanlu} describe a tool CFix that can detect and fix
concurrency bugs by identifying bug patterns in the code.
CFix also simplifies its own patches by merging fixes for related bugs.

Usui et al.~\cite{UBES10} provide a dynamic adaptive lock placement
algorithm, which can be precise but necessitates runtime overhead.

Our abstraction is based on the one from~\cite{CAV15}, which is
similar to abstractions that track reads and writes to individual locations
(e.g.,~\cite{VYRS10,AKNP14}). 
In~\cite{VYY10} the authors rely on
assertions for synchronization synthesis and
include iterative abstraction refinement in their framework, which
could be integrated in our approach. 
%Mention related work with some optimization criteria (maximal
%permissiveness [VYY09], quantitative synthesis)  

\subsection{Illustrative Example}

One of the main contributions of the paper is a synthesis approach
that allows optimization for a particular computing platform and program. 
 We will demonstrate on an example that varying
parameters of our performance model, which corresponds to varying the
machine architecture and usage pattern (contention), can lead to a different
solution with best performance. 

\begin{figure}[t]
\includegraphics[trim = 0.0in 0.0in 0.0in
  0.0in,clip,width=0.65\linewidth]{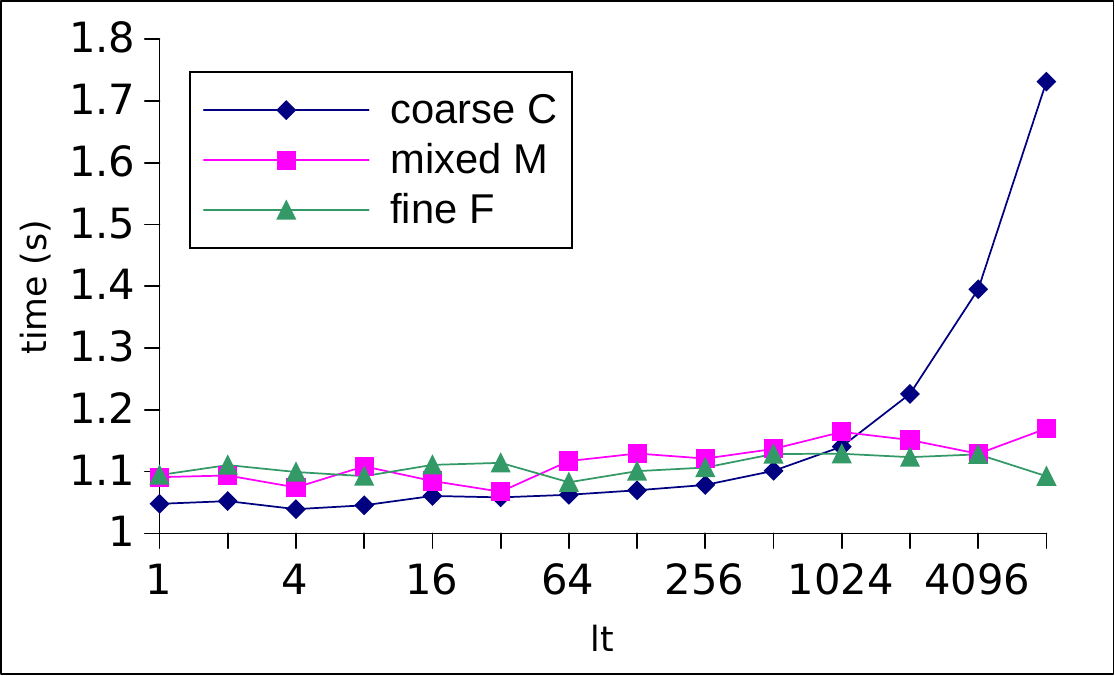}  
\caption{Performance under various locking schemes}
\label{fig:pgraph}
\end{figure}

Consider the function {\tt worker\_thread} in
Figure~\ref{ex:worksharing} that is called by a number
of worker threads in a work-sharing setting. The  {\tt worker\_thread}
function calls three functions that access shared memory: {\tt
  sharedX()}, {\tt sharedY()}, and {\tt sharedZ()}. Each of the three
functions needs to be called mutually exclusively with itself, but it
does not conflict with the other two functions. None of the three
functions uses locks internally. 
The function {\tt local()} does not access shared memory. 
If we consider locking schemes that use only one lock, and lock only
within an iteration of 
the loop, then there are a number of correct locking schemes. First, there
is a coarse-grained version $C$ which locks before the
call to {\tt sharedX()} and unlocks after the call to {\tt
  sharedZ()}. Second, there is a fine-grained version $F$ that locks each
function separately. Third, there are
intermediate versions, such as a version $M$ that locks
before the call to {\tt sharedX()}, unlocks before the call to {\tt
  local()}, and then it locks again the call to {\tt sharedZ()}. 

Clearly, which of the correct versions will have the best performance
depends on the program, the contention,
and the architecture. For instance, with very low contention, the
coarse-grained version would be fastest. On the other hand, with
contention, and if {\tt local()} is expensive 
enough, the third version performs best, as it releases the lock
before calling {\tt local()}.  

We further demonstrate with a small experiment (Figure~\ref{fig:pgraph}) 
that often no locking scheme is uniformly
better. We considered variants of the program in
Figure~\ref{ex:worksharing} which differ in how long it takes to
execute the function {\tt local()} (parameter lt in
Figure~\ref{fig:pgraph}). We kept the other parameters (such as
contention and number of threads) constant (there were $4$
threads). We see that if the call to {\tt local} is cheap, then the
coarse-grained version performs better, but it can be perform very
badly otherwise. The other two versions $M$ and $F$ perform
comparably. 

The example is similar to our main case study of the memcached
server. We demonstrate on that case study that the performance model
helps us choose the option with the best performance, and that it does not
have to be the most coarse-grained or the most fine-grained locking
solution.

\section{Formal Framework}
\label{sec:defs}

We present the syntax and semantics of a {\em concrete} concurrent while
language $\wlang$, followed by the syntax and semantics of an {\em abstract}
concurrent while language $\wlangabs$.  While $\wlang$ (and our tool) permits
function call and return statements, we skip these constructs in the
formalization below. We conclude the section by formalizing our notion of 
correctness for concrete concurrent programs. \\

In our work, we assume a read or a write to a single shared variable executes
atomically and further assume a sequentially consistent memory model. 

\subsection{Concurrent Programs}

\noindent{\bf\em Syntax of $\wlang$ (\figref{wlang_syntax}).}
A concurrent program $\cProg$ is a finite collection of threads $\langle
\thread_1, \ldots, \thread_n \rangle$ where each thread is a statement written
in the syntax of $\wlang$. 
All $\wlang$ variables (shared program variables \texttt{s\_var}, local program
variables  \texttt{l\_var}, lock variables \texttt{lk\_var},
condition variables \texttt{c\_var} and guard variables \guard)  range over
integers. Each statement is labeled with a unique location identifier ${\tt
loc}$; we denote by $\stmt{{\tt loc}}$ the statement labeled by ${\tt loc}$. 
%\ttsays{What about local variables? They could exist in the concrete program (you mention them in a footnote even)}

The language $\wlang$ includes standard syntactic constructs such as
assignment, conditional, loop, synchronization, {\tt goto} and {\tt yield}
statements. In $\wlang$, we only permit expressions that read from at most one
shared variable and assignments that either read from or write to exactly one
shared variable\footnote{An expression/assignment statement that involves
reading from/writing to multiple shared variables can always be rewritten into
a sequence of atomic read/atomic write statements using local variables.}. 
The language also includes {\tt assume}, {\tt assume\_not}, {\tt set} and {\tt
unset} statements whose use will be clarified later. 
Most significantly, $\wlang$ permits reading from ($\Input{\sf tag}$) and
writing to ($\Output{\sf tag}{s\_expr}$) a communication channel {\tt tag} between
the program and an interface to an external system. 
%We assume that the program and the external system interface 
%can only communicate through the channel. 
%and No program variable is visible to the external system.  
In practice, we use the tags to model device registers.  
In our presentation, we consider only a single external interface. 

\begin{figure}

%\vspace{1mm}
\hbra
%\vspace{-2mm}
\begin{verbatim}
l_expr::= constant | * | l_var | 
          operator(l_expr, l_expr, ... , l_expr)
s_expr::= s_var | 
          operator(s_var, l_expr, ..., l_expr)
lstmt ::= loc: stmt | lstmt; lstmt
stmt  ::= s_var := l_expr | l_var := s_expr |
          s_var := havoc() | while (s_expr) lstmt | 
          if (s_expr) lstmt else lstmt | 
          s_var := in(tag) | out(tag,s_expr) | 
          lock(lk_var)  | unlock(lk_var) | 
          wait(c_var) | wait_not(c_var) | 
          notify(c_var) | reset(c_var) 
          assume(g_var) | assume_not(g_var) | 
          set(g_var) | unset(g_var) | 
          goto loc | yield | skip
\end{verbatim}
%\ttsays{Why can s\_expr only operate on one variable and constants. It should operate also on local variables.}
\vspace{-2mm}
\hket
%\vspace{-2mm}
\caption{Syntax of $\wlang$ \label{fig:wlang_syntax}}
\end{figure}

\noindent{\bf\em Semantics of $\wlang$.}
We first define the semantics of a single thread in $\wlang$, 
and then extend the definition to concurrent non-preemptive and preemptive semantics. 

\noindent{\em Single-thread semantics (\figref{single_thread_semantics}).}
Let us fix a thread identifier $\tid$. We use $\tid$ interchangeably with the program it represents. 
A state of a single thread is given by $\langle \VariableValuation, \loc \rangle$ where $\VariableValuation$
is a valuation of all program variables visible to thread $\tid$, and $\loc$ is a location 
identifier, indicating the statement in $\tid$ to be executed next. 

We define the {\em flow graph} $\cfg_{\tid}$ for thread $\tid$ in a manner
similar to the control-flow graph of $\tid$. Each node of $\cfg_{\tid}$ is
labeled with a unique labeled statement of $\tid$ (unlike a control-flow graph,
statements in the same basic block are not merged into a single node). The
edges of $\cfg_{\tid}$ capture the flow of control in $\tid$.  Nodes labeled
with $\mtt{if(s\_expr)}$ and $\mtt{while(s\_expr)}$ statements have two
outgoing edges, labeled with $\mtt{assume \; s\_expr}$ and $\mtt{assume \; \neg
s\_expr}$, respectively. The flow graph $\cfg_{\tid}$ has a unique entry node and a unique exit node. 
The entry node is the first labeled statement in $\tid$; we denote its location 
identifier by $\mtt{first}_{\tid}$. The exit node is a special node 
corresponding to a hypothetical statement $\mtt{last}_{\tid}:\, \mtt{skip}$ placed at the end of
$\tid$.  
%\ttsays{A CFG has the statements in the nodes, the edges are labeled
%with assumes or nothing}\\ 
%if $\ell: \mtt{if (s\_expr) \; \ell_{1}: stmt_1 \; else \; \ell_{2}: stmt_2}$
%is a statement in $\tid$, then node $\ell$ in $\cfg_{\tid}$ has outgoing edges
%to nodes $\ell_{1}$ and $\ell_{2}$, labeled with $\mtt{assume \; s\_expr}$ and
%$\mtt{assume \; \neg s\_expr}$, respectively.  

We define successors of locations of $\tid$ using $\cfg_{\tid}$. 
The location \last~has no successors. 
We define $\suc(\ell) = \ell'$ if node $\ell: \mathtt{stmt}$ in $\cfg_{\tid}$ 
has exactly one outgoing edge to node $\ell': \mathtt{stmt}'$.
We define $\suc_1(\ell) = \ell_1$ and $\suc_2(\ell) = \ell_2$ if 
node $\ell: \mathtt{stmt}$ in $\cfg_{\tid}$  
has exactly two outgoing edges to nodes $\ell_1: \mathtt{stmt}_1$ and 
$\ell_2: \mathtt{stmt}_2$.
% We use the notation $\evalExpression{\e}$ to denote the value of
% expression $\e$ under the valuation $\VariableValuation$, and
% $\VariableValuation[x := k]$ to denote the valuation obtained from
% $\VariableValuation$ by replacing the value assigned to variable $x$
% with $k$.

We can now define the single-thread operational semantics.
A single execution step $\smallstepx{\VariableValuation,
\loc}{\VariableValuation', \loc'}{\alpha}$ changes the program state
from $\langle\VariableValuation, \loc\rangle$ to $\langle\VariableValuation', \loc'\rangle$, while
optionally outputting an {\em observable symbol} $\alpha$. 
The absence of a symbol is denoted using $\epsilon$.
In the following, $\e$ represents an expression and $\evalExpression{\e}$ evaluates an
expression by replacing all variables $v$ with their values in $\VariableValuation$.

In \autoref{fig:single_thread_semantics}, we present a partial set of 
rules for single execution steps. The only rules which 
involve output of an observable are:
%{\sc Havoc}, {\sc Input}, and {\sc Output} rules:
% We have that:
\begin{compactenum}
\item {\sc Havoc}: Statement $\loc: \Asgn{v}{\Havoc}$ assigns $v$ a non-deterministic
  value (say $k$) and outputs the observable $(\tid, \Havoc, k, v)$.
\item {\sc Input}, {\sc Output}: $\loc: \Asgn{x}{\Input{t}}$ and $\loc: \Output{t}{\e}$
  read and write values to the channel $t$, and output $(\tid, \mathtt{in}, k, t)$
  and $(\tid, \mathtt{out}, k, t)$, where $k$ is the value read
  or written, respectively.
\end{compactenum}
Intuitively, the observables record the sequence of non-deterministic
guesses, as well as the input/output interaction with the tagged
channels. The semantics of the synchronization statements shown in 
\figref{single_thread_semantics} is standard. 

The semantics of {\tt assume}, {\tt assume\_not}, {\tt set} and {\tt unset} statements 
are identical to that of {\tt wait}, {\tt wait\_not}, {\tt notify} and {\tt reset} statements, 
respectively. Thus, $\mathtt{assume}(g)$ and $\mathtt{assume\_not}(g)$ 
execute iff guard variable $g$ equals $1$ and $0$, respectively. Statements 
$\mathtt{set}(g)$ and $\mathtt{unset}(g)$ assign $1$ and $0$ to guard variable $g$, respectively.
%\rs{Do we need the complete rules in the Appendix?}

\begin{figure}
  \caption{A partial set of rules for single-thread semantics of $\wlang$\label{fig:single_thread_semantics}}
  %\begin{display}{}
  \hbra
    \staterule{Havoc}
    {\stmt{\loc} = \Asgn{v}{\Havoc} \qquad k \in \natset \qquad \loc' = \suc(\loc) }
    {\smallstepx{\VariableValuation,\loc}{\VariableValuation[x:=k],\loc'}{(\tid, \Havoc, k, x)}} 
    %{\smallstepo{\loc:\ \Asgn{x}{\Havoc}}{[x:=k],\suc(\loc)}{\alpha}}% {\suc(\loc)}{\alpha}{x:=k}}
    \\[\GAP]
    \staterule{Input}
    {\stmt{\loc} = \Asgn{v}{\Input{t}} \qquad k \in \natset \qquad \loc' = \suc(\loc)}
    {\smallstepx{\VariableValuation,\loc}{\VariableValuation[x:=k],\loc'}{(\tid,{\tt in}, k,t)}} 
    %{\smallstepo{\loc:\ \Asgn{x}{\Havoc}}{[x:=k],\suc(\loc)}{\alpha}}% {\suc(\loc)}{\alpha}{x:=k}}
    \\[\GAP]
    \staterule{Output}
    {\stmt{\loc} = \Output{t}{\e}  \quad \evalExpression{\e} = k \quad \loc' = \suc(\loc)}
    {\smallstepx{\VariableValuation,\loc}{\VariableValuation,\loc'}{(\tid,{\tt \out}, k,t)}} 
    \\[\GAP]
    \staterule{Lock}
    {\stmt{\loc} = \mathtt{lock}(lk) \qquad \VariableValuation[lk] = 0 \qquad \loc' = \suc(\loc)}
    {\smallstepx{\VariableValuation,\loc}{\VariableValuation[lk := \tid],\loc'}{\epsilon}}
    \\[\GAP]
    \staterule{Unlock}
    {\stmt{\loc} = \mathtt{unlock}(lk) \qquad \VariableValuation[lk] = \tid \qquad \loc' = \suc(\loc)}
    {\smallstepx{\VariableValuation,\loc}{\VariableValuation[lk := \tid],\loc'}{\epsilon}}
    \\[\GAP]
    \staterule{Wait}
    {\stmt{\loc} = \mathtt{wait}(cv) \qquad \VariableValuation[cv] = 1 \qquad \loc' = \suc(\loc)}
    {\smallstepx{\VariableValuation,\loc}{\VariableValuation,\loc'}{\epsilon}}
    \\[\GAP]
    \staterule{Wait\_not}
    {\stmt{\loc} = \mathtt{wait\_not}(cv) \quad \VariableValuation[cv] = 0 \quad \loc' = \suc(\loc)}
    {\smallstepx{\VariableValuation,\loc}{\VariableValuation,\loc'}{\epsilon}}
    \\[\GAP]
    \staterule{Notify/Reset}
    {\stmt{\loc} = \mathtt{notify}(cv)/\mathtt{reset}(cv) \qquad \loc' = \suc(\loc)}
    {\smallstepx{\VariableValuation,\loc}{\VariableValuation[cv:=1/0],\loc'}{\epsilon}}
    \\[\GAP]
%    \staterule{Assume}
%    {\stmt{\loc} = \mathtt{assume}(g) \qquad g = 1 \qquad \loc' = \suc(\loc)}
%    {\smallstepx{\VariableValuation,\loc}{\VariableValuation,\loc'}{\epsilon}}
%    \\[\GAP]
%    \staterule{Assume\_not}
%    {\stmt{\loc} = \mathtt{assume\_not}(g) \quad g = 0 \quad \loc' = \suc(\loc)}
%    {\smallstepx{\VariableValuation,\loc}{\VariableValuation,\loc'}{\epsilon}}
%    \\[\GAP]
%\staterule{Set/Unset}
%    {\stmt{\loc} = \mathtt{set}(g)/\mathtt{unset}(g) \qquad \loc' = \suc(\loc)}
%    {\smallstepx{\VariableValuation,\loc}{\VariableValuation[g:=1/0],\loc'}{\epsilon}}
%    \\[\GAP]
\vspace{-2mm}
\hket
%\end{display}
%\ttsays{These semantics allow double locking by the same thread. Is this intentional?}\\
%\rs{I kept the same semantics as in the CAV15 paper. Have now removed double locking.}
\vspace{-2mm}
\end{figure}

\noindent{\em Concurrent semantics.}
A state of a concurrent program is given by $\langle \VariableValuation, ctid,
(\loc_1, \ldots, \loc_n) \rangle$ where $\VariableValuation$ is a valuation of
all program variables, $ctid$ is the thread identifier of the currently
executing thread and $\loc_1, \ldots, \loc_n$ are the locations of the
statements to be executed next in threads $\thread_1$ to $\thread_n$,
respectively. Initially, all program variables and $ctid$ equal $0$ and for
each $i \in [1,n]: \loc_i = \mathtt{first}_i$.  
%For each lock variable $lk$,
%$\VariableValuation[lk]$ equals $0$ if no thread holds the lock and $\tid$ if
%thread $\tid$ currently holds the lock. For each condition variable $c$,
%$\VariableValuation[c]$ equals $1$ after $\mathtt{notify}(c)$
%executes and $0$ after $\mathtt{reset}(c)$ executes.  
%\com{May not need the above. Should be part of semantics.}
%
\begin{figure}
  \caption{Non-preemptive semantics\label{fig:nonpreemptive_semantics}}
  \begin{display}{}
    \staterule{Seq}
    {ctid = i \qquad \smallstepx{\VariableValuation , \ell_i}{\VariableValuation', \ell_i'}{\alpha}}
    {\smallstepx{\VariableValuation,   ctid, (\ldots, \loc_i, \ldots)}{\VariableValuation,
      ctid, (\ldots, \loc_i', \ldots)}{\alpha}}
    \\[\GAP]
    
    \staterule{Thread\_end}
    {ctid = i \quad  \loc_i = \mathtt{last}_i \quad  ctid' \in \{1,\ldots,n\}\setminus\{i\}}
    {\smallstep{\VariableValuation, ctid, (\ldots, \loc_i, \ldots)}
                {\VariableValuation, ctid', (\ldots, \loc_i, \ldots)}}
    \\[\GAP]
    \vspace{0.7ex}

    \staterule{Nswitch}
    {\stmt{\loc_i} = \mathtt{lock}(lk)/\mathtt{wait}(cv)/\mathtt{wait\_not}(cv)/\mathtt{yield}\\[\GAP] 
    ctid = i \qquad  ctid' \in \{1,\ldots,n\} \qquad \loc_i' = \suc(\loc)}
    {\smallstep{\VariableValuation, ctid, (\ldots, \loc_i, \ldots)}
                {\VariableValuation, ctid', (\ldots, \loc_i', \ldots)}}
  \end{display}
  %\ttsays{Some rules duplicate what the single thread semantics already has (signal, reset)}
\vspace{-2mm}
\end{figure}

\begin{figure}
  \caption{Additional rule for preemptive semantics\label{fig:preemptive_semantics}}
  \begin{display}{}
    \staterule{Pswitch}
    {ctid' \in \{1,\ldots,n\}}
    {\smallstep{\VariableValuation, ctid, (\loc_1, \ldots, \loc_n)}
               {\VariableValuation, ctid', (\loc_1, \ldots, \loc_n)}}
  \end{display}
\vspace{-2mm}
\end{figure}

\noindent{\em Non-preemptive semantics (\figref{nonpreemptive_semantics}).} 
The non-preemptive semantics ensures that a single thread from the
program keeps executing using the single-thread semantics (Rule {\sc Seq}) until one of the following
occurs:
\begin{inparaenum}[(a)]
\item the thread finishes execution (Rule {\sc Thread\_end}) or it encounters a 
\item {\tt yield}, {\tt lock},  {\tt wait} or {\tt wait\_not} statement (Rule {\sc Nswitch}).
\end{inparaenum}
In these cases, a context-switch is possible. \\
%\rs{I did not emphasize the unconditional yielding at locks}
%\ttsays{Well what you state is wrong. Await can switch if the condition is met. This is our semantics in the implementation.}

\noindent{\em Preemptive semantics (\figref{nonpreemptive_semantics}, \figref{preemptive_semantics}).}
The preemptive semantics of a program is obtained from the
non-preemptive semantics by relaxing the condition on context-switches,
and allowing context-switches at all program points.
In particular, the preemptive semantics consist of the rules of the non-preemptive semantics
and the single rule {\sc Pswitch} in \figref{preemptive_semantics}.
\begin{figure}[tb]
%\scriptsize{
%\begin{alltt}
%int open = 0; // device interface void power\_up(); void power\_down(); 
%\end{alltt}
%} 
\begin{minipage}{0.45\textwidth}%
%\scriptsize{
\begin{alltt}
void open\_dev() \{
1: while (*) \{
2:   if (open==0) 
3:      power\_up();
4:   open:=open+1;
5: yield; \} \}
\end{alltt}
%}
\end{minipage}
\vrule
\hspace{5mm}
\begin{minipage}{0.45\textwidth}%
%\scriptsize{
\begin{alltt}
void open\_dev\_abs() \{
1: while (*) \{
2: r(open);
   if (*) 
3:    w(dev);
4: r(open);
   w(open);
5: yield; \} \}
\end{alltt}
%}
\end{minipage}
\vspace{-2ex}
\caption{Example procedure and its abstraction}
\label{fig:example1}
\end{figure}

\subsection{Abstract Concurrent Programs}\label{sec:abstraction}
For concurrent programs written in $\wlang$ communicating with external system
interfaces, it suffices to focus on a simple, data-oblivious abstraction (\cite{CAV15}). 
The abstraction tracks types of accesses (read or write) to each memory location
while abstracting away their values. 
Inputs/outputs to an external interface 
are modeled as writes to a special memory location ({\tt dev}).
%Even inputs are modeled as writes because in our applications we cannot assume that reads from
%the external interface are free of side-effects. Havocs become
%ordinary writes to the variable they are assigned to.
Havocs become
ordinary writes to the variable they are assigned to.
Every branch is taken non-deterministically and tracked. 
Given $\cProg$ written in $\wlang$, we denote by $\cProgabs$ 
the corresponding abstract program written in $\wlangabs$. 

\noindent {\em Example.} We present a procedure {\tt open\_dev()} and its
abstraction in \figref{example1}. The function {\tt power\_up()} represents a
call to a device.

%\ttsays{Should we give a formal abstraction in the appendix? How to get from a concrete program to its abstract program.}\\
%\rs{I thought about this, but decided the above description is enough. Let's revisit this later.}

\noindent{\bf\em Abstract Syntax (\figref{wlangabs_syntax}).}
In the figure, {\tt var} denotes all shared program variables 
and the {\tt dev} variable. Observe that the abstraction 
respects the valuations of the lock, condition and guard variables\footnote{The purpose of the 
guard variables is to improve the precision of our otherwise coarse abstraction. Currently, 
they are inferred manually, but can presumably be inferred automatically 
using an iterative abstraction-refinement loop. In our current benchmarks, guard variables 
needed to be introduced in only three scenarios.}. 
 
\begin{figure}

%\vspace{1mm}
\hbra
%\vspace{-2mm}
\begin{verbatim}
lstmt ::= loc: stmt | lstmt; lstmt
stmt  ::= r(var) | w(var) | if(*) lstmt else lstmt  
          | while(*) lstmt | 
          lock(lk_var)  | unlock(lk_var) | 
          wait(c_var) | wait_not(c_var) |
          notify(c_var) | reset(c_var) |
          assume(g_var) | assume_not(g_var) |
          set(g_var) | unset(g_var) | 
          goto loc | yield | skip
\end{verbatim}
\vspace{-2mm}
\hket
%\vspace{-2mm}
\caption{Syntax of $\wlangabs$ \label{fig:wlangabs_syntax}}
\end{figure}

\noindent{\bf\em Abstract Semantics.}
As before, we first define the semantics of $\wlangabs$ for a single-thread.

\noindent{\em Single-thread semantics (\figref{abstract_semantics}.)} The abstract state of a single thread
$\tid$ is given simply by $\langle \loc \rangle$ where $\loc$ is the location
of the statement in $\tid$ to be executed next. 
We define the flow graph and successors for locations in the abstract program 
$\tid$ in the same way as before.
An abstract observable symbol is of the form: $(\tid, \theta,\ell)$, where 
$\theta \in \{\mathtt{(read, v),(write, v),if,else,loop,exitloop}\}$. 
The symbol $\theta$ records the type of access to variables along with the variable name 
 $(\mathtt{(read, v),(write, v)})$ and records non-deterministic branching choices 
$\{\mathtt{if,else,loop,exitloop}\}$. 
\figref{abstract_semantics} presents the rules for statements unique to 
$\wlangabs$; the rules for statements common to $\wlangabs$ and $\wlang$ are the same. 

%\ttsays{This is dangerous. The observables have to have the information which thread they come from (better even which location). Otherwise a read from one thread could be matched by the language inclusion to a read from another thread.}\\
%
\begin{figure}
  \caption{Partial set of rules for single-thread semantics of $\wlangabs$\label{fig:abstract_semantics}}
  \begin{display}{}
    \staterule{Read}
    {\stmt{\loc} = \mathtt{r}(v) \qquad \loc' = \suc(\loc)}
    {\smallstepx{\loc}{\loc'}{(\tid, (\mathtt{read}, v),\loc)}}
    \\[\GAP]
    \staterule{Write}
    {\stmt{\loc} = \mathtt{w}(v) \qquad \loc' = \suc(\loc)}
    {\smallstepx{\loc}{\loc'}{(\tid, (\mathtt{write}, v),\loc)}}
    \\[\GAP]
    \staterule{If1}
    {\stmt{\loc} = \mathtt{if}(*) \; s_1 \; \mathtt{else} \; s_2 \qquad \loc' = \suc_1(\loc)}
    {\smallstepx{\loc}{\loc'}{(\tid, \mathtt{if}, \loc)}}
    \\[\GAP]
    \staterule{If2}
    {\stmt{\loc} = \mathtt{if}(*) \; s_1 \; \mathtt{else} \; s_2 \qquad \loc' = \suc_2(\loc)}
    {\smallstepx{\loc}{\loc'}{(\tid, \mathtt{else}, \loc)}}
    \\[\GAP]
    \staterule{While1}
    {\stmt{\loc} = \mathtt{while}(*) \; s \qquad \loc' = \suc_1(\loc)}
    {\smallstepx{\loc}{\loc'}{(\tid, \mathtt{loop}, \loc)}}
    \\[\GAP]
    \staterule{While2}
    {\stmt{\loc} = \mathtt{while}(*) \; s \qquad \loc' = \suc_2(\loc)}
    {\smallstepx{\loc}{\loc'}{(\tid, \mathtt{exitloop}, \loc)}}
  \end{display}
%  \vspace{-4ex}
\end{figure}

\noindent{\em Concurrent semantics.} A state of an abstract concurrent program
is given by $\langle \VariableValuation_o, ctid, (\loc_1, \ldots, \loc_n)
\rangle$ where $\VariableValuation_o$ is a valuation of all lock, condition and
guard variables, $ctid$ is the current thread identifier and $\loc_1, \ldots,
\loc_n$ are the locations of the statements to be executed next in threads
$\thread_1$ to $\thread_n$, respectively. The non-preemptive and preemptive
semantics of a concurrent program written in $\wlangabs$ are defined in the same way as that
of a concurrent program written in $\wlang$. 

\subsection{Executions and Observable Behaviours}

Let $\wprog$, $\wprogabs$ denote the set of all concurrent programs in $\wlang$, $\wlangabs$, 
respectively. 

\noindent{\bf\em Executions.} A {\em non-preemptive/preemptive execution} of a concurrent
program $\cProg$ in $\wprog$ is an alternating sequence of program states and
(possibly empty) observable symbols, $\ProgState_0 \alpha_1 \ProgState_1 \ldots
\alpha_k \ProgState_k$, such that (a) $\ProgState_0$ is the initial state of $\cProg$
and (b) $\forall j \in [0,k-1]$, according to the
non-preemptive/preemptive semantics of $\wlang$, we have
$\ProgState_j \xrightarrow{\alpha_{j+1}}{\ProgState_{j+1}}$. 
%(c) $\loc_1,\ldots,\loc_n = \mtt{last}_{1},\ldots,\mtt{last}_{n}$ in $\ProgState_k$.  
A non-preemptive/preemptive execution of a concurrent program $\cProgabs$ in $\wprogabs$ 
is defined in the same way, replacing the corresponding semantics of $\wlang$ with that of $\wlangabs$.

Given an execution $\pi$, let $\mathsf{obs}(\pi)$ denote the sequence of
non-empty observable symbols in $\pi$.
 
\noindent{\bf\em Observable Behaviours.}
The {\em non-preemptive}/{\em preemptive observable behaviour} of program
$\cProg$ in $\wprog$, denoted $\sem{\cProg}^{NP}$/$\sem{\cProg}^{P}$, is the
set of all sequences $\omega$ of non-empty observable symbols such that $\omega = \mathsf{obs}(\pi)$ for 
some non-preemptive/preemptive execution $\pi$ of $\cProg$.  
The {\em non-preemptive}/{\em preemptive observable behaviour} of program
$\cProgabs$ in $\wprogabs$, denoted $\sem{\cProgabs}^{NP}$/$\sem{\cProgabs}^{P}$, 
is defined similarly.

\subsection{Program Correctness}

We specify correctness of concurrent programs in $\wprog$ using three {\em implicit}
criteria, presented below. 

\noindent{\bf\em Preemption-safety.}
Observable behaviours $\omega_1$ and $\omega_2$ of a program $\cProg$ in $\wprog$ are 
{\em equivalent} if: 
\begin{inparaenum}[(a)]
\item the subsequences of $\omega_1$ and
  $\omega_2$ containing only symbols of the form
  $(\tid, \mathtt{in}, k, t)$ and $(\tid, \mathtt{out}, k,
  t)$ are equal and
\item for each thread identifier $\tid$, the subsequences of
  $\omega_1$ and $\omega_2$ containing only
  symbols of the form $(\tid, \mathtt{havoc}, k, x)$ are equal.
\end{inparaenum}
Intuitively, observable behaviours are equivalent if they have the same
interaction with the interface, and the same non-deterministic choices
in each thread.
For sets $\mathcal{O}_1$ and $\mathcal{O}_2$ of observable behaviours, we
write $\mathcal{O}_1 \subseteq \mathcal{O}_2$ to denote that each sequence
in $\mathcal{O}_1$ has an equivalent sequence in
$\mathcal{O}_2$.

%
%Formally, a {\em non-preemptive observation sequences} of $\cProg$ is a sequence 
%$\alpha_0\ldots\alpha_k$ such that there exist program states
%$\ProgState_0^{pre}$, $\ProgState_0^{post}$, \ldots,
%% $\ProgState_1^{pre}$, $\ProgState_1^{post}$, 
%$\ProgState_k^{pre}$, $\ProgState_k^{post}$ such that according to the
%non-preemptive semantics of $\wlang$, we have:
%\begin{inparaenum}[(a)]
%\item for each $0 \leq i \leq k$,
%  $\smallstepx{\ProgState_i^{pre}}{\ProgState_i^{post}}{\alpha_i}$,
%\item for each $0 \leq i < k$,
%  $\smallstepxs{\ProgState_i^{post}}{\ProgState_{i+1}^{pre}}{\epsilon}$,
%  and
%\item for the initial state $\ProgState_\iota$ and a final state (i.e., where
%  all threads have finished execution) $\ProgState_f$,
%  $\smallstepxs{\ProgState_{\iota}}{\ProgState_0^{pre}}{\epsilon}$ and
%  $\smallstepxs{\ProgState_k^{post}}{\ProgState_{f}}{\epsilon}$.
%\end{inparaenum}
%Similarly, a {\em preemptive observation sequence} of a program
%$\cProg$ is a sequence
%$\alpha_0\ldots\alpha_k$ as above, with the non-preemptive semantics
%replaced with preemptive semantics.
%We denote the sets of non-preemptive and preemptive observation
%sequences of a program $\cProg$ by $\sem{\cProg}^{NP}$ and
%$\sem{\cProg}^P$, respectively.
%
Given a concurrent programs $\cProg$ and $\cProg'$ in $\wprog$ such that $\cProg'$ is 
obtained by adding locks to $\cProg$, $\cProg'$ is {\em preemption-safe} w.r.t. $\cProg$ if 
$\sem{\cProg'}^{P} \subseteq \sem{\cProg}^{NP}$.

\noindent{\bf\em Deadlock-freedom.} 
A state $\ProgState$ of concurrent program $\cProg$ in $\wprog$ is a 
{\em deadlock state} under non-preemptive/preemptive semantics if  
\begin{compactenum}[(a)]
\item there exists a non-preemptive/preemptive execution %$\ProgState_0 \alpha_1 \ProgState_1 \ldots
%\alpha_k \ProgState$ 
from the initial state $\ProgState_0$ of $\cProg$ to $\ProgState$,
%, such that $\ProgState_0$ is the initial state of $\cProg$ 
%and $\forall j \in [0,k-1]$, according to the non-preemptive semantics of $\wlang$, 
%$\smallstepx{\ProgState_j}{\ProgState_{j+1}}{\alpha_{j+1}}$, 

\item there exists thread $i$ such that $\loc_i \neq \mtt{last}_i$ in $\ProgState$, and 

\item $\neg \exists \ProgState'$: $\smallstepx{\ProgState}{\ProgState'}{\alpha}$ according to the 
nonpreemptive/preemptive semantics of $\wlang$. 
\end{compactenum}
Program $\cProg$ in $\wprog$ is {\em deadlock-free 
under non-preemptive/preemptive semantics} 
if no non-preemptive/preemptive execution of $\cProg$ hits a
deadlock state.  In other words, every
non-preemptive/preemptive execution of $\cProg$ ends in a state with
$\loc_1,\ldots,\loc_n = \mtt{last}_{1},\ldots,\mtt{last}_{n}$.
We say $\cProg$ is {\em deadlock-free} if it is deadlock-free under 
both non-preemptive and preemptive semantics. 
%\rs{Did not mention the assume statements}
%No thread progresses (no enabled successor) and\\
%$\exists tid: loc_{tid} \neq last \ \wedge \ stmt(loc_{tid}) \neq \mtt{assume}$\\ 

%Every thread terminates\\

%Can traces be infinite?\\

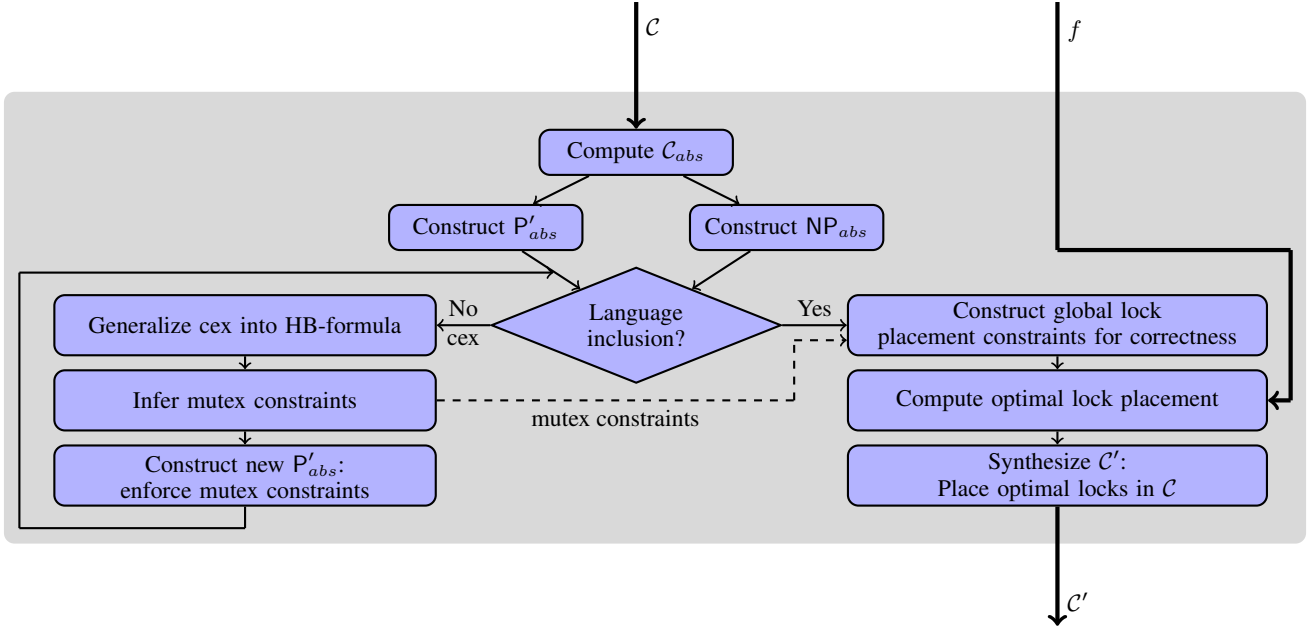
\begin{figure*}[t!]
\begin{minipage}{0.99\linewidth}
\begin{tikzpicture}
\pgfdeclarelayer{background}

\tikzstyle{decision} = [draw,diamond,aspect=2.5,fill=blue!30, thick,
    text=black, text width=17mm,text centered]
%[diamond, draw, fill=blue!30, thick, 
  %  text width=50mm, text badly centered, node distance=25mm, inner sep=0pt]
\tikzstyle{sblock} = [draw,rectangle,rounded corners,fill=blue!30, thick, minimum
    width=7mm, minimum height=6mm, text width=25mm,
    inner sep=1pt,text=black,text centered]
\tikzstyle{nblock} = [draw,rectangle,rounded corners,fill=blue!30, thick, minimum
    width=7mm, minimum height=8mm, text width=50mm,
    inner sep=1pt,text=black,text centered]
\tikzstyle{block} = [draw,rectangle,rounded corners,fill=blue!30, thick, minimum
    width=7mm, minimum height=8mm, text width=55mm,
    inner sep=1pt,text=black,text centered]

\node [sblock] (init) {Compute $\cProgabs$};
%
%\node [sblock, below of=init, node distance=8mm] (init2) {Set $\cProgabs' =\cProgabs$};
%
\node [sblock, below of=init, node distance=10mm, xshift=2cm] (np) 
{Construct $\autNP$};
\node [sblock, below of=init, node distance=10mm, xshift=-2cm] (p) 
{Construct $\autP$};
\node [decision, below of=init, node distance=23mm] (lang) 
{Language inclusion?};
%{$\sem{\cProgabs'}^{P} \subseteq \sem{\cProgabs}^{NP}$?};
%
\node [nblock, left of=lang, node distance=52mm] (gen) {Generalize cex into \hbformula};
\node [nblock, below of=gen, node distance=10mm] (mutex) {Infer mutex constraints};
\node [nblock, below of=mutex, node distance=10mm] (prune) {Construct new $\autP$:\\ 
enforce mutex constraints};

\node [block, right of=lang, node distance=56mm] (smt) {Construct global lock \\ placement constraints for correctness};% to ensure deadlock-freedom and legitimate locking discipline};
\node [block, below of=smt, node distance=10mm] (opt) {Compute optimal lock placement};
\node [block, below of=opt, node distance=10mm] (out) {Synthesize $\cProg'$:\\Place optimal locks in $\cProg$};
\node[coordinate, node distance=28mm, left of=smt, yshift=-2mm] (c0) {};
\node[coordinate, node distance=35mm, left of=smt, yshift=-2mm] (c1) {};
\node[coordinate, node distance=8mm, below of=c1] (c2) {};

\node[coordinate, node distance=7mm, below of=prune] (d0) {};
\node[coordinate, node distance=30mm, left of=d0] (d1) {};
\node[coordinate, node distance=34mm, above of=d1] (d2) {};
\node[coordinate, node distance=71mm, right of=d2] (d3) {};
\node[coordinate, node distance=43mm, above of=smt] (e0) {};
\node[coordinate, node distance=33mm, below of=e0] (e1) {};
\node[coordinate, node distance=31mm, right of=e1] (e2) {};
\node[coordinate, node distance=31mm, right of=opt] (e3) {};
\node[coordinate, node distance=28mm, right of=opt] (e4) {};

\node[coordinate, node distance=20mm, above of=init] (i) {};
\node[coordinate, node distance=20mm, below of=out] (end) {};

\node[coordinate, node distance=7mm, above of=init] (top) {};
\node[coordinate, node distance=8mm, below of=out] (bottom) {};
\node[coordinate, node distance=1mm, left of=d1] (left) {};
\node[coordinate, node distance=4mm, right of=e4] (right) {};
%
%\path[->] (init) edge (init2);
\path[thick,->] (init) edge (np);
\path[thick,->] (init) edge (p);
\path[thick,->] (np) edge (lang);
\path[thick,->] (p) edge (lang);

\path[thick,->] (lang) edge node[above]{Yes}  (smt);
\path[thick,->] (lang) edge node[above]{No} node[below]{cex} (gen);
\path[thick,->] (gen) edge (mutex);
\path[thick,->] (mutex) edge (prune);

\path[thick,->] (smt) edge (opt);
\path[thick,->] (opt) edge (out);

\path[thick,dashed,-] (mutex) edge node[below]{mutex constraints} (c2);
\path[thick,dashed,-] (c2) edge (c1);
\path[thick,dashed,->] (c1) edge (c0);

\path[thick,-] (prune) edge (d0);
\path[thick,-] (d0) edge (d1);
\path[thick,-] (d1) edge (d2);
\path[thick,->] (d2) edge (d3);

\path[ultra thick,->] (e3) edge (e4);
\path[ultra thick,-] (e3) edge (e2);
\path[ultra thick,-] (e2) edge (e1);
\path[ultra thick,-] (e0) edge node[above right,pos=0.2]{$f$}(e1);

\path[ultra thick,->] (i) edge node[right, pos=0.2]{$\cProg$} (init);

\path[ultra thick,->] (out) edge node[right, pos=0.8]{$\cProg'$}(end);

\begin{pgfonlayer}{background}
\node[rectangle, fill=gray!30, rounded corners, fit= (top) (bottom) (left) (right), label=above:{}]{};
\end{pgfonlayer}

\end{tikzpicture}
\end{minipage}
\caption{Solution Overview \label{fig:overview}}
\end{figure*}

\noindent{\bf\em Legitimacy of locking discipline.}
Let us first fix some notation for execution steps of a concurrent program 
$\cProg$ in $\wprog$:
\begin{compactitem}
\item Let $\mathtt{lock}_{\tid,lk}$ denote the single step execution of a 
$\mathtt{lock}(lk)$ statement in thread $\tid$:\\ 
$\smallstepx{(\VariableValuation, \tid, (\ldots, \loc_{\tid}, \ldots)}{\VariableValuation',
      \tid, (\ldots, \suc(\loc_{\tid}), \ldots)}{\epsilon}$ 
where 
$\stmt{\loc_{\tid}} = \mathtt{lock}(lk)$, $\VariableValuation[lk] = 0$ and $\VariableValuation'[lk] = \tid$. 
\item Similarly, let $\mathtt{unlock}_{\tid,lk}$ denote the single step execution of an 
$\mathtt{unlock}(lk)$ statement in thread $\tid$.
%$\smallstepx{(\VariableValuation, \tid, (\ldots, \loc_{\tid}, \ldots)}{\VariableValuation',
%      \tid, (\ldots, \suc(\loc_{\tid}), \ldots)}{\epsilon}$ 
%where 
%$\stmt{\loc_{\tid}} = \mathtt{unlock}(lk)$, $\VariableValuation[lk] = \tid$ and $\VariableValuation'[lk] = 0$. 
\item Given an execution $\pi = \ProgState_0 \alpha_1 \ProgState_1 \ldots \alpha_k \ProgState_k$ of $\cProg$, 
let $\pi[\xrightarrow{j}]$ denote the single execution step 
$\ProgState_{j} \xrightarrow{\alpha_{j+1}} \ProgState_{j+1}$ in $\pi$.
\end{compactitem}
Program $\cProg$ has {\em legitimate locking discipline under non-preemptive/\\
preemptive semantics} if for any nonpreemptive/preemptive execution $\pi$ of $\cProg$, the 
following are true:
\begin{compactenum}[(a)]
\item {\em Lock implies eventually (but not immediately after) unlock}:\\
$\exists i$: $\pi[\xrightarrow{i}] = \mathtt{lock}_{\tid,lk}$ 
$\Rightarrow$ $\exists j > i+1$: $\pi[\xrightarrow{j}] = \mathtt{unlock}_{\tid,lk}$ 
%\\ \rs{Do we want this to be $j > i + 1$?}
\item {\em Unlock implies earlier (but not immediately before) lock}:\\ 
$\exists i$: $\pi[\xrightarrow{i}] = \mathtt{unlock}_{\tid,lk}$ 
$\Rightarrow$ $\exists j < i-1$: $\pi[\xrightarrow{j}] = \mathtt{lock}_{\tid,lk}$
\item {\em No double locking}:\\ 
$\exists i,j$: $i<j$, $\pi[\xrightarrow{i}] = \mathtt{lock}_{\tid,lk}$ and 
$\pi[\xrightarrow{j}] = \mathtt{lock}_{\tid,lk}$\\ 
$\Rightarrow$ $\exists m$: $i+1 < m < j-1$ and $\pi[\xrightarrow{m}] = \mathtt{unlock}_{\tid,lk}$
\item {\em No double unlocking}:\\ 
$\exists i,j$: $i<j$, $\pi[\xrightarrow{i}] = \mathtt{unlock}_{\tid,lk}$ and 
$\pi[\xrightarrow{j}] = \mathtt{unlock}_{\tid,lk}$\\
$\Rightarrow$ $\exists m$: $i+1 < m < j-1$ and $\pi[\xrightarrow{m}] = \mathtt{lock}_{\tid,lk}$
\end{compactenum}
We say $\cProg$ has {\em legitimate locking discipline} if it has 
legitimate locking discipline under both non-preemptive and preemptive semantics. 
semantics
%\rs{Implementation: No locks in immutable parts of the program}

\section{Problem Statement and Solution Overview}
\label{sec:new}

\subsection{Problem Statement}

Given a concurrent program $\cProg$ in $\wprog$ such that $\cProg$ is 
deadlock-free and has legitimate locking discipline under non-preemptive semantics, 
{\em and} an objective function $f:\wprog \mapsto \Reals$, the goal is to 
synthesize a new concurrent program $\cProg'$ in $\wprog$ such that:
\begin{compactenum}[(a)]
\item $\cProg'$ is obtained by adding locks to $\cProg$, 
\item $\cProg'$ is preemption-safe w.r.t. $\cProg$,
\item $\cProg'$ is deadlock-free, 
\item $\cProg'$ has legitimate locking discipline, {\em and},
\item $\cProg' = \underset{\cProg'\in \wprog \text{satisfying (a)-(d) above}}{\arg\,\min} \; f$ 
\end{compactenum}

\subsection{Solution Overview}

Our solution framework (\figref{overview}) consists of the following main components.

\noindent{\bf\em Reduction of preemption-safety to language inclusion \cite{CAV15}.}
To ensure tractability of checking preemption-safety, we rely on the
abstraction described in \secref{abstraction}.  Observable behaviours
$\omega_1$ and $\omega_2$ of an abstract program $\cProgabs$ in $\wprogabs$
are {\em equivalent} if (a) they are equal \upto~the classical independence
relation $I$ on memory accesses: accesses to different locations are
independent, and accesses to the same location are independent iff they are
both read accesses and (b) subsequences of $\omega_1$ and $\omega_2$ with 
symbols $(\tid, \theta, \loc)$, $\theta \in \{\mathtt{if,else,loop,exitloop}\}$, 
are equal. Using this notion of equivalence, the notion of
preemption-safety is extended to abstract programs.

Under abstraction, we model each thread as a nondeterministic finite
automaton (\nfa) over a finite alphabet consisting of abstract observable symbols. 
This enables us to construct \nfas\ $\autNP$ and $\autP$ accepting the languages 
$\sem{\cProgabs}^{NP}$ and $\sem{\cProgabs'}^{P}$, respectively ($\cProgabs$ is 
the abstract program corresponding to $\cProg$ and initially, $\cProgabs' = \cProgabs$). 
It turns out that preemption-safety of $\cProg'$ 
w.r.t. $\cProg$ is implied by preemption-safety of $\cProgabs'$ w.r.t. $\cProgabs$, 
which, in turn, is implied by {\em language inclusion \upto~$I$} of \nfas~$\autP$ and $\autNP$. 
\nfas~$\autP$ and $\autNP$ satisfy language inclusion \upto~$I$ if any word accepted by 
$\autP$ is equivalent to some word obtainable by repeatedly commuting 
adjacent independent symbol pairs in a word accepted by $\autNP$. 
While the problem of language inclusion \upto~an independence relation
is undecidable~\cite{bertoni1982equivalence}, we 
define and decide a bounded version of language inclusion \upto~an independence relation\footnote{Our language inclusion procedure starts with an initial bound and iteratively increases the bound until it
reports that the inclusion holds, or finds a counterexample, or reaches a timeout}.

\noindent{\bf\em Inference of mutex constraints from generalized counterexamples.}
If $\autP$ and $\autNP$ do not satisfy language inclusion \upto~$I$, then we 
obtain a counterexample $cex$ and analyze it to infer constraints on 
$\Lang{\autP}$ for eliminating $cex$. Our counterexample analysis 
examines the set $nhood(cex)$ of all permutations of the symbols 
in $cex$ that are accepted by $\autP$. The output of the counterexample analysis 
is an {\em hbformula} $\phi$ --- a Boolean combination of {\em happens-before} 
or {ordering} constraints between events --- representing all counterexamples 
in $nhood(cex)$. Thus $cex$ is generalized into a larger set of counterexamples 
represented as $\phi$.

From $\phi$, we infer possible locks-enforceable constraints on $\Lang{\autP}$
that can eliminate all counterexamples satisfying $\phi$. 
The key observation we exploit is that common concurrency bugs manifest as 
simple patterns of ordering constraints between events. For instance, the pattern 
$(\tid_1,\theta_1,\loc_1) < (tid_2,\theta_2,\loc_2)  \; \wedge \; 
(\tid_2,\theta'_1,\loc'_2) < (\tid_1,\theta'_2,\loc'_1)$, indicates an atomicity violation 
and can be rewritten as a {\em mutual exclusion (mutex) constraint}: $\text{mutex}(\tid_1.[\loc_1:\loc'_1],\tid_2.[\loc_2:\loc'_2])$. Note that this mutex constraint can be easily enforced by a lock 
that protects access to the regions $\loc_1$ to $\loc'_1$ in $\tid_1$ and 
$\loc_2$ to $\loc'_2$ in $\tid_2$. We refer the reader to \cite{POPL15} for more details.  

\noindent{\bf\em Automaton modification for enforcing mutex constraints.}
Once we have the mutex constraints inferred from a generalized counterexample, 
we can insert the corresponding locks into $\cProgabs'$, reconstruct $\autP$ 
and repeat the process, starting from checking $\autP$ and $\autNP$ for language inclusion modulo $I$. 
This is a greedy iterative loop for synchronization synthesis and is undesirable (see \secref{intro}). 
Hence, instead of modifying $\cProgabs'$ in each iteration by inserting locks, 
we modify $\autP$ in each iteration to enforce the mutex constraints on $\Lang{\autP}$ 
and then repeat the process.  
We describe the procedure for modifying $\autP$ to enforce mutex constraints in \secref{automod}.

\noindent{\bf\em Construction of global lock placement constraints.}
Once $\autP$ and $\autNP$ satisfy language inclusion modulo $I$, we formulate 
global constraints over lock placements for ensuring correctness. 
These global constraints include all mutex constraints inferred over all 
iterations and constraints for enforcing deadlock-freedom and 
legitimacy of lock placement. Any model of the global constraints 
corresponds to a lock placement that ensures program correctness. 
We describe the formulation of these global constraints in \secref{globalconstraints}.

\noindent{\bf\em Computation of $f$-optimal lock placement.} 
In our final component, given an objective function $f$, we compute 
a lock placement that satisfies the global constraints and is $f$-optimal. 
We then synthesize the final output $\cProg'$ by inserting the computed 
lock placement in $\cProg$. We present various objective functions 
and describe the computation of their respective optimal solutions in \secref{quantitative}.

\section{Enforcing Mutex Constraints in $\boldsymbol{\autP}$}
\label{sec:automod}
To enforce mutex constraints in $\autP$, we prune paths in
$\autP$ that violate the mutex constraints. 
%\rs{Explain $\loc_1:\loc_1'$}\\ 

\noindent{\bf\em Conflicts.} Given a mutex constraint
$\text{mutex}(\tid_1.[\loc_1:\loc'_1],\tid_2.[\loc_2:\loc'_2])$, 
a {\em conflict} is a tuple 
$(\loc_i^{\pre},\loc_i^{\mi},\loc_i^{\post},\loc_j^{\conpre},\loc_j^{\conpost})$ 
of location identifiers satisfying the following:
\begin{inparaenum}[(a)] 
\item $\loc_i^{\pre}$, $\loc_i^{\mi}$, $\loc_i^{\post}$ are adjacent locations  
 in thread $\tid_i$ for $i \in [1,2]$,
\item $\loc_j^{\conpre}$, $\loc_j^{\conpost}$  are adjacent locations in the other thread $\tid_{3-i}$, 
\item $\loc_i \leq \loc_i^{\pre}, \loc_i^{\mi}, \loc_i^{\post} \leq \loc_i'$ and
\item $\loc_j \leq \loc_j^{\conpre}, \loc_j^{\conpost} \leq \loc_j'$. 
\end{inparaenum}
Intuitively, a conflict represents a {\em minimal violation} of a mutex constraint 
due to the execution of a statement in thread $j$ between two adjacent statements in thread $i$. 
The execution of each statement is represented using a source location and a destination location. 

Given a conflict $c = (\loc_i^{\pre},\loc_i^{\mi},\loc_i^{\post},\loc_j^{\conpre},\loc_j^{\conpost})$,  
let $\pre(c) = \loc_i^{\pre}$, $\mi(c) = \loc_i^{\mi}$, $\post(c) = \loc_i^{\post}$, 
$\conpre(c) = \loc_j^{\conpre}$ and $\conpost(c) = \loc_j^{\conpost}$. Further, let 
$\tid_1(c) = i$ and $\tid_2(c) = j$.  
Let $\Conflicts$ denote the set of all conflicts derived
from all mutex constraints in the current loop iteration 
and let $\n = |\Conflicts|$. 

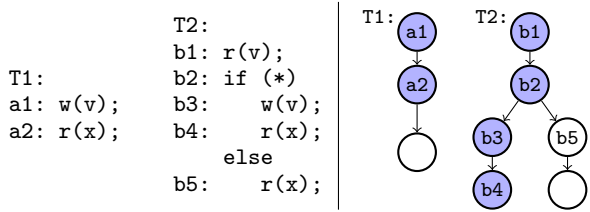
\begin{figure}[tb]
%\scriptsize{
%\begin{alltt}
%int open = 0; // device interface void power\_up(); void power\_down(); 
%\end{alltt}
%} 
\begin{minipage}{0.25\textwidth}%
%\scriptsize{
\begin{alltt}
T1:
a1: w(v);
a2: r(x);
\end{alltt}
%}
\end{minipage}
\begin{minipage}{0.25\textwidth}%
%\scriptsize{
\begin{alltt}
T2:
b1: r(v);
b2: if (*) 
b3:    w(v);
b4:    r(x);
    else 
b5:    r(x);
\end{alltt}
%}
\end{minipage}
\vrule
\hspace{2mm}
\begin{minipage}{0.3\textwidth}
\begin{tikzpicture}
\small{
    \tikzstyle{state}=[draw,circle,minimum size=5mm,thick,inner sep=1pt,text=black,fill=blue!30]
    \tikzstyle{lbl}=[font=\fontsize{9}{9}\selectfont,inner
        sep=1pt,minimum height=2mm]
    \tikzstyle{statelab}=[minimum size=4mm,thick,inner sep=1pt,text=black]
    
    \node[state]              (a) {{\tt a1}};
    \node[state, node distance=7mm, below of=a] (b) {{\tt a2}};
    \node[state, fill=none, node distance=9mm, below of=b] (c) {}; %{{\tt last}};

    \node[state, node distance=15mm, right of=a] (A) {{\tt b1}};
    \node[state, node distance=7mm, below of=A] (F) {{\tt b2}};
    \node[state, node distance=7mm, below of=F,xshift=-5mm] (B) {{\tt b3}};
    \node[state, fill=none, node distance=7mm, below of=F,xshift=5mm] (C) {{\tt b5}};
    %\node[state, node distance=14mm, below of=F] (D) {{\tt last}};
    \node[state, node distance=7mm, below of=B] (D) {{\tt b4}};
    \node[state, fill=none, node distance=7mm, below of=C] (E) {};
    
    \node[statelab,node distance=5mm,left of=a,yshift=2mm]  () {{\tt T1:}};
    \node[statelab,node distance=5mm,left of=A,yshift=2mm]  () {{\tt T2:}};

    \path[->] (a) edge (b);
    \path[->] (b) edge (c);

    \path[->] (A) edge (F);
    \path[->] (F) edge (B);
    \path[->] (F) edge (C);
    \path[->] (B) edge (D);
    \path[->] (C) edge (E);
}
\end{tikzpicture}
\end{minipage}
\vspace{-2ex}
\caption{Example: Mutex constraints and conflicts}
\label{fig:ex_conflict}
\end{figure}

\noindent{\em Example.} We have an example program and its flow-graph in
\figref{ex_conflict} (we skip the statement labels in the nodes here). Suppose
in some iteration we obtain $\text{mutex}(\mtt{T1.[a1:a2],T2.[b1:last]})$. This
yields 2 conflicts: $c_1$ given by $(\mtt{b1,b2,b3,a1,a2})$ and $c_2$ given by
$(\mtt{b2,b3,b4,a1,a2})$. On an aside, this example also illustrates the
inadequacy of a greedy approach for lock placement.  The mutex constraint
yields a lock $\mtt{lock}(\mtt{T1.[a1:a2],T2.[b1:last]})$.  This is not a
legitimate lock placement; in executions executing the {\tt else} branch, the
lock is never released. 

\noindent{\bf\em Constructing new $\autP$.}
Initially, \nfa~$\autP$ is given by the tuple
$(S_{\old},\Sigma \cup \{\epsilon\},\ProgState_{0,\old},F_{\old},T_{\old})$, 
where
\begin{inparaenum}[(a)]
\item $S_{\old}$ is the set of states $\langle \VariableValuation, ctid, (\loc_1,\ldots,\loc_n)\rangle$
of the abstract program $\cProgabs$ corresponding to $\cProg$, 
\item $\Sigma$ is the set of abstract observable symbols, 
\item $\ProgState_{0,\old}$ is the initial state of $\cProgabs$, 
\item $F_{\old}$ is the set of states in $\cProgabs$ with $\loc_1,\ldots,\loc_n = \mtt{last}_{1},\ldots,\mtt{last}_{n}$ and 
\item $T_{\old} \subseteq S \times \Sigma \times S$ is the transition relation with 
$(\ProgState, \alpha, \ProgState) \in T$ iff $\ProgState \xrightarrow{\alpha} \ProgState'$ 
according to the abstract preemptive semantics. 
\end{inparaenum}

To enable pruning paths that violate mutex constraints, we augment the state
space of $\autP$ to track the status of conflicts $c_1,\ldots, c_\n$ using {\em
four-valued} propositions $p_1,\ldots, p_\n$, respectively.  Initially all
propositions are $0$. Proposition $p_k$ is incremented from $0$ to $1$ when
conflict $c_k$ is {\em activated}, i.e., when control moves from
$\loc_i^{\pre}$ to $\loc_i^{\mi}$ along a path. Proposition $p_k$ is
incremented from $1$ to $2$ when conflict $c_k$ {\em progresses}, i.e., when thread
$\tid_i$ is at $\loc_i^{\mi}$ and control moves from $\loc_j^{\conpre}$ to
$\loc_j^{\conpost}$. Proposition $p_k$ is incremented from $2$ to $3$ when
conflict $c_k$ {\em completes}, i.e., when control moves from $\loc_i^{\mi}$ to
$\loc_i^{\post}$. Proposition $p_k$ is reset to $0$ when conflict $c_k$ is
{\em aborted}, i.e., when thread $\tid_i$ is at $\loc_i^{\mi}$ and either moves to a
location different from $\loc_i^{\post}$, or moves to $\loc_i^{\post}$ before
thread $\tid_j$ moves from $\loc_j^{\conpre}$ to $\loc_j^{\conpost}$.

\noindent{\em Example.} In \figref{ex_conflict}, $c_1$ is activated 
when $\mtt{T2}$ moves from $\mtt{b1}$ to $\mtt{b2}$; $c_1$ progresses 
if now $\mtt{T1}$ moves from $\mtt{a1}$ to $\mtt{a2}$ and is aborted
if instead $\mtt{T2}$ moves from $\mtt{b2}$ to $\mtt{b3}$; $c_2$ completes 
after progressing if $\mtt{T2}$ moves from $\mtt{b2}$ to $\mtt{b3}$ and is aborted 
if instead $\mtt{T2}$ moves from $\mtt{b2}$ to $\mtt{b5}$.

Formally, the new $\autP$ is given by the tuple  
$(S_{\new},\Sigma \cup \{\epsilon\},s_{0,new},F_{\new},T_{\new})$, where:
\begin{compactenum}[(a)]
\item $S_{\new} = S_{\old} \times \{0,1,2,3\}^\Conflicts$,
\item $\Sigma$ is the set of abstract observable symbols as before, 
\item $\ProgState_{0,\new} = (\ProgState_{0,\old},(0,\ldots,0))$,
\item $F_{\new} = \{(\ProgState, (0,\ldots,0)): \ProgState \in F_{\old}\}$ and
\item $T_{\new}$ is constructed as follows: \\
add $((\ProgState,(p_1,\ldots,p_\n)),\alpha, (\ProgState',(p'_1,\ldots,p'_\n)))$ to $T_{\new}$ iff \\
$(\ProgState, \alpha, \ProgState') \in T_{\old}$ and for each $k \in [1,\n]$, the following hold:
\begin{compactenum}[1.]
\item {\em Conflict activation:} \\
if $p_k = 0$, $ctid = ctid' = \tid_1(c_k)$, $\loc_{ctid} = \pre(c_k)$ and $\loc'_{ctid} = \mi(c_k)$, 
then $p'_k = 1$,\\
\item {\em Conflict progress:}\\ 
else if $p_k = 1$, $ctid = ctid' = \tid_2(c_k)$, $\loc_{\tid_1(c_k)} = \loc'_{\tid_1(c_k)} = \mi(c_k)$,  
$\loc_{ctid} = \conpre(c_k)$ and $\loc'_{ctid} = \conpost(c_k)$,  
then $p'_k = 2$, 
\item {\em Conflict completion and state pruning:}\\
else if $p_k = 2$, $ctid = ctid' = \tid_1(c_k)$,  
$\loc_{ctid} = \mi(c_k)$ and $\loc'_{ctid} = \post(c_k)$,  
then $p'_k = 3$; 
delete state $(\ProgState',(p'_1,\ldots,p'_\n))$\footnote{Our mutex constraints 
are in disjunctive normal form (CNF). %,  and hence our conflicts are in conjunctive normal form. 
Hence, in our implementation, we track conflict completions and delete a state 
only after the DNF is falsified.},
\item {\em Conflict abortion - $\tid_1(c_k)$ executes alternate statement:}\\ 
else if $p_k = 1$ or $2$, $ctid = ctid' = \tid_1(c_k)$,   
$\loc_{ctid} = \mi(c_k)$ and $\loc'_{ctid} \neq \post(c_k)$,  
then $p'_k = 0$,
\item {\em Conflict abortion - $\tid_1(c_k)$ executes before $tid_2(c_k)$:}\\ 
else if $p_k = 1$, $ctid = ctid' = \tid_1(c_k)$, $\loc_{ctid} = \mi(c_k)$ and $\loc'_{ctid} = \post(c_k)$, 
$\loc_{\tid_2(c_k)} = \loc'_{\tid_2(c_k)} = \conpre(c_k)$,  
then $p'_k = 0$
\end{compactenum}
\end{compactenum}

In our implementation, the new $\autP$ is constructed on-the-fly. Moreover, we
do not maintain the entire set of propositions $p_1,\ldots,p_\n$ in each state
of $\autP$. A proposition $p_i$ is added to the list of tracked propositions
only after conflict $c_i$ is activated. Once conflict $c_i$ is aborted, $p_i$
is dropped from the list of tracked propositions. 

%\begin{itemize}
%\item {\em Conflict progress:}\\ 
%else if $p_k = 1$, $\ProgState = \langle \VariableValuation, ctid, 
%(\ldots,\post(p_k),\ldots,\conflict(p_k),\ldots)\rangle$ 
%and $\ProgState' = \langle \VariableValuation', ctid, 
%(\ldots,\post(c_k),\ldots,\suc(\conflict(c_k)),\ldots)\rangle$, then
%$c'_k = 2$,  
%\item For each $k \in [1,\n]$:  
%if $c_k = 2$, $\ProgState = \langle \VariableValuation, ctid, 
%(\ldots,\post(c_k),\ldots)\rangle$ 
%and $\ProgState' = \langle \VariableValuation', ctid, 
%(\ldots,\suc(\post(c_k)),\ldots)\rangle$, then
%$c'_k = 3$,  
%%\item For each $k \in [1,\n]$: 
%%if $c_k = 1$, $\ProgState = \langle \VariableValuation, ctid, 
%%(\ldots,\post(c_k),\ldots,\conflict(c_k),\ldots)\rangle$, 
%%$\ProgState' \neq \langle \VariableValuation, ctid, (\ldots,\suc(\post(c_k)),\ldots)\rangle$ and 
%%$\ProgState' \neq \langle \VariableValuation, ctid, (\ldots,\suc(\post(c_k)),\ldots)\rangle$ and 
%%\item Assuming the past statement is from thread A. If any other statement than next is executed or next is executed before the conflict statement is executed than the conflict-id is removed from the list.

\section{Global Lock Placement Constraints}
\label{sec:globalconstraints}

We encode the global lock placement constraints for ensuring correctness 
as an SMT\footnote{The encoding of the global lock placement constraints is 
essentially a SAT formula. We present and use this as an SMT formula to enable combining 
the encoding with objective functions for optimization (see \secref{quantitative}).
\ttsays{The objective functions i considered where all in SAT}} 
formula $\smt$. Let $\Loc$ denote the set of all 
location and $\Locks$ denote the set of all locks available 
for synthesis. We use scalars $\x, \x', \x_1, \ldots$ of type $\Loc$  
to denote locations and scalars $\lk, \lk', \lk_1, \ldots$ of type 
$\Locks$ to denote locks. Let $\pred(\x)$ denote the set of all 
predecessors in node $\x: \stmt{\x}$ in the flow-graph of 
the current abstract concurrent program $\cProgabs'$. Let 
$\AllConflicts$ denote the set of all 
conflicts derived from mutex constraints across all iterations.
We use the following Boolean variables in the encoding.\\ 
\begin{tabular}{l l}
\hline
$\LockBefore(\x,\lk)$ & $\mtt{lock}(\lk)$ is placed just before $\x$\\
$\LockAfter(\x,\lk)$ & $\mtt{lock}(\lk)$ is placed just after $\x: \stmt{\x}$\\
$\UnlockBefore(\x,\lk)$ & $\mtt{unlock}(\lk)$ is placed just before $\x$\\
$\UnlockAfter(\x,\lk)$ & $\mtt{unlock}(\lk)$ is placed just after $\x: \stmt{\x}$\\
$\ord(\lk,\lk')$ & when nesting $\lk$, $\lk'$,\\
&                  $\mtt{lock}(\lk)$ is placed before $\mtt{lock}(\lk')$\\
\hline
\end{tabular}
%We use the following macros in the encoding.\\
%\begin{tabular}{l l l}
%\hline
%$\InLock(\x,\lk)$ & $\x$ is protected by $\lk$ \\
%\multicolumn{2}{l}
%{
%$[\LockBefore(\x,\lk) \vee (\neg \UnlockBefore(\x,\lk) \wedge \underset{\x' \in \pred(x)}\bigvee \InLockExt(\x',\lk))]$}\\ 
%$\InLockExt(\x,\lk)$ & \rs{??} \\
%\multicolumn{2}{l} 
%{
%$[(\InLock(\x,\lk) \wedge \neg \UnlockAfter(\x,\lk)) \vee \LockAfter(\x,\lk)]$
%}\\
%\hline
%\end{tabular}
\vspace{2mm}

We describe the main constraints constituting $\smt$ below. For illustrative purposes, 
we also present the SMT formulation for some of these constraints. 
All constraints are over each $\x \in \Loc$ and $\lk \in \Locks$.
\begin{compactenum}
\item Define $\InLock(\x,\lk)$: $\x$ is protected by $\lk$.\\
$\InLock(\x,\lk) = \LockBefore(\x,\lk)\ \vee\ (\neg \UnlockBefore(\x,\lk)\; \wedge$\\ 
$\underset{\x' \in \pred(x)}\bigvee \InLockExt(\x',\lk))$ 
\item Define $\InLockExt(\x',\lk)$: $\x$ is protected by $\lk$ or $\mtt{lock}(\lk)$ is placed after $\x$.\\
$(\InLock(\x,\lk) \wedge \neg \UnlockAfter(\x,\lk)) \vee \LockAfter(\x,\lk)$
\item All locations in the same conflict in $\AllConflicts$ 
are protected by the same lock, without interruption. 
\item Placing $\mtt{lock}(\lk)$ immediately before/after $\mtt{unlock}(\lk)$ is disallowed.\\
$\UnlockBefore(\x,\lk) \Rightarrow (\neg \underset{\x' \in \pred(x)}\bigvee \LockAfter(\x',\lk))$\\
$\LockBefore(\x,\lk) \Rightarrow (\neg \underset{\x' \in \pred(x)}\bigvee \UnlockAfter(\x',\lk))$
\item Enforce the lock order.
\item No {\tt wait} statements in the scope of synthesized locks. \\
$(\underset{cv}\bigvee \stmt{\x} = \mtt{wait}(cv)) \Rightarrow \neg \InLock(\x,\lk)$
\item Placing both $\mtt{lock}(\lk)$ and $\mtt{unlock}(\lk)$ before/after $\x$ is disallowed. \\
%\underset{\x \in \Loc, \lk \in \Locks}\bigwedge 
$(\neg \LockBefore(\x,\lk) \vee \neg \UnlockBefore(\x,\lk))\; \wedge$\\ 
$(\neg \LockAfter(\x,\lk) \vee \neg \UnlockAfter(\x,\lk))$
%end{align*}
\item All predecessors must agree on their $\InLockExt$ status.\\ 
$(\underset{\x' \in \pred(x)}\bigwedge \ \InLockExt(\x',\lk))\; \vee$\\ 
$(\underset{\x' \in \pred(x)}\bigwedge \ \neg \InLockExt(\x',\lk))$
\item $\mtt{unlock}(\lk)$ can be placed only after $\mtt{lock}(\lk)$ has been placed.\\
$\UnlockAfter(\x,\lk) \Rightarrow \InLock(\x,\lk)$\\
$\UnlockBefore(\x,\lk) \Rightarrow \underset{\x' \in \pred(x)}\bigvee\ \InLockExt(\x,\lk)$
\item No double locking.
\item No double unlocking.
\end{compactenum}

Besides the above, $\smt$ includes constraints that account for existing locks in the
program.  In particular, the lock order enforces all synthesized locks to be
placed after any existing lock (when nesting). Further, $\smt$ 
also includes constraints enforcing uniformity of lock placements in 
loop bodies.

We have the following result.

\begin{theorem}
Let concurrent program $\cProg'$ be obtained by inserting any lock placement satisfying $\smt$ 
into concurrent program $\cProg$. Then $\cProg'$ is guaranteed to be preemption-safe w.r.t. $\cProg$, 
deadlock-free and have legitimate locking discipline.
\end{theorem}

%In the above, constraint 1 corresponds to preemption-safety, constraints 2-3
%correspond to deadlock-freedom and constraint 5-8 correspond to legitimacy of
%locking discipline. All other constraints ensure that the encoding 

\section{Optimizing Lock Placement}
\label{sec:quantitative}

The global lock placement constraint $\smt$ constructed in
Section~\ref{sec:globalconstraints} often has multiple models
corresponding to very different lock placements.
The desirability of these lock placements vary considerably due to
performance considerations.
Hence, any comprehensive approach to synchronization synthesis needs to take into account
performance while generating the solution program.

Here, we present two types of objective functions to
distinguish between different lock placements, as well as
accompanying optimization procedures.
These procedures take as input the global lock placement constraints
along with any auxiliary inputs required by an objective function 
$f$ and produce the lock placement that ensures program correctness 
and is $f$-optimal. 

The first category of objective functions consists of formal encodings of
various syntactic ``rules of thumb'' (such as minimality of critical sections)
used by programmers, while the second category is based on building a
performance model based on profiling.  The second category is less widely
applicable than the first, but also corresponds more closely to real
performance on a machine.

\subsection{Syntactic optimization}
\label{sec:syntactic_optimization}
%In this section, we present three optimization criteria that are 
%are all based on syntactic rules for lock placement and do not 
%use any performance profiling.  
%based on rules of thumb used by concurrent programmers.
%These criteria share the commonality that they are based on syntactic
%placement of locks rather than any measurable performance criteria.  

We say that a statement $\instr$ in a concurrent program $\cProg$ is
protected by a lock $lk$ if $\InLock(\instr,\lk)$ is true. 
We define two syntactic objective functions as follows:
\begin{enumerate}
%  \item {\em Minimal critical sections.} 
%    Under this optimization criterion, a program $\cProg_1$ is considered
%    better than $\cProg_2$ if the total number of statements that are
%    protected by any lock in $\cProg_1$ is smaller than in $\cProg_2$.
%    Formally, we can define $\AbsMin(\cProg_i)$ to be the count of
%    statements in $\cProg_i$ that are protected by any lock.
  \item {\em Coarsest locking.}
    Under this objective function, a program $\cProg_1$ is considered
    better than $\cProg_2$ if the number of lock statements in
    $\cProg_1$ is fewer than in $\cProg_2$. 
    Among the programs having the same number of lock statements, the
    ones with the fewest statements protected by any lock are
    considered better.
    Formally, we can define $\Coarsest(\cProg_i)$ to be $\lambda +
    \epsilon \cdot \AbsMin(\cProg_i)$ where $\lambda$ is the count of 
    {\tt lock} statements  in $\cProg_i$, $\AbsMin(\cProg_i)$ is the count of
    statements in $\cProg_i$ that are protected by any lock and $\epsilon$ is given by
    $\frac{1}{2k}$ where $k$ is the total number of statements in $\cProg_i$.
  \item {\em Finest locking.}
    This objective function asks for maximum possible concurrency,
    i.e., it asks to minimize the number of pairs of statements from
    different threads that cannot be executed together.
    Formally, we define $\Finest(\cProg_i)$ to be the sum over pairs
    of statements $\instr_1$ and $\instr_2$ from different threads
    that cannot be executed at the same time, i.e., are protected by the
    same lock.
\end{enumerate}

\noindent{\bf\em A note on usage.}
Neither of the objective functions mentioned above are guaranteed to
find the optimally performing program in all scenarios.
It is necessary for the programmer to judge when each criterion is to be
used.
Intuitively, coarsest locks should be used when the cost of locking
operations is relatively high compared to th cost of executing the
critical sections, while the finest locks should be used when locking
operations are cheap compared to the cost of executing the critical
sections.
%The finest locks ensure maximal concurrency as defined
%in~\cite{dontrememberthepaper}.
%The minimal critical sections criterion is often used as a guiding
%criterion for automated tools (see, for example,~\cite{xxx,yyy,zzz}).
%Further, the critical sections generated by the minimal critical sections
%criterion can be replaced by atomic sections to obtain a program with
%atomic sections.
%This program is optimal in terms of the size of the atomic sections, and can
%be executed in execution models that allow for atomic sections (e.g.,
%transactional memory).

\noindent{\bf\em Optimization procedure.}
The main idea behind the optimization procedure for the above syntactic
objective functions is to build an instance of the MaxSMT problem using the
global lock placement constraint $\smt$, such that
\begin{inparaenum}[(a)]
  \item Every model of $\smt$ is a model for the MaxSMT problem; and
  \item the cost of each model for the MaxSMT problem is the cost of the
    corresponding locking scheme according to the chosen objective function.
\end{inparaenum}
The optimal lock placement is then computed by solving the MaxSMT problem.

A MaxSMT problem instance is given by $\langle \Phi,
\langle (\Psi_1, w_1), \ldots \rangle \rangle$ where $\Phi$ and
$\Psi_i$'s are SMT formulae and the $c_i$'s are real numbers.
The formula $\Phi$ is called the {\em hard constraint}, and each
$\Psi_i$ is called a {\em soft constraint} with $w_i$ being its 
{\em associated weight}.
Given an assignment $V$ of variables occurring in the constraints, its
cost is defined to be $\sum_{i | V \not \models \Psi_i} w_i$.
The objective of the MaxSMT problem is to find a model that
satisfies $\Phi$ while having the minimal cost.

In the following, we write $\InLock(\instr)$ as a short-hand
for $\bigvee_{lk} \InLock(\instr, lk)$.
For each of the above objective functions, the hard constraint for the
MaxSMT problem is $\smt$ and the soft constraints and associated
weights are as specified below:
\begin{compactitem}
%\item For the minimal critical sections criterion, the soft constraints
%  $\Psi_i$ are given by $\neg \InLock(\instr)$ for each
%  statement $\instr$ in the program.
%  The associated cost of each $\Psi_i$ is $1$.
\item For the coarsest locking objective function, the soft constraints are of
  three types: 
  \begin{inparaenum}[(a)]
  \item $\LockBefore(\instr)$ with weight $1$,
  \item $\LockAfter(\instr)$ with weight $1$, and 
  \item $\neg \InLock(\instr)$ with weight $\epsilon$, where $\epsilon$ is as defined above.
  \end{inparaenum}
\item For the finest locking objective function, the soft constraints are given by
  $\bigwedge_{lk} \neg \InLock(\instr, lk) \vee
  \neg \InLock(\instr', lk)$, for each pair of statements $\instr$ and
  $\instr'$ from different threads.
  The weight of each soft constraint is $1$.
\end{compactitem}

%It is easy to see that in each of the above MaxSMT problems, the optimal
%model has the same cost as the corresponding lock placement according
%to respective criteria.
\begin{theorem}
  For each of the three syntactic objective functions, the cost of the optimal
  program is equal to the cost of the model for the
  corresponding MaxSMT problem obtained as described above.
\end{theorem}

\subsection{Performance profiling-based optimization}
\label{sec:profiling_optimization}
\rs{Change "c" notation}
We first present the performance model for a very restricted setting.
The target use case of this performance model is for high-performance
work-sharing code as present in servers or database management systems.
In such systems, it is usually the case that there are as many worker
threads as there are cores in the CPU with each thread executing the
same code. 
We build the performance model based on this execution model.
We also assume that the synthesizer is allowed to introduce at most one
lock variable, which may be taken and released multiple times by each
thread during its execution, i.e., the various programs the synthesizer
can synthesize differ only in the granularity of locking.
We later show how this assumption of only one lock variable can be
removed.

\noindent{\bf\em Execution model.}
To formally define the parameters that our performance model depends on, we
first need to understand and define the characteristics of the execution model.
The ``average'' time taken to execute a concurrent program depends on a wide
variety of factors (e.g., underlying hardware platform, distribution of inputs,
and scheduler).
Here, we list a number of assumptions we make about these factors:
\begin{compactitem}
\item {\em Usage and scheduling models.}
  We assume that under intended usage of the program, program inputs
  are drawn from a probability distribution and further, that the
  expected running time of the program is finite.
  We also assume that the system scheduler is {\em oblivious}, i.e.,
  does not make scheduling choices based on the actual values of
  variables in the program.
  This is a reasonable assumption for most systems.
  Note that neither the input probability distribution, nor the system
  scheduler needs to be formally modelled or known--all we require is
  the ability to obtain ``typical'' inputs and run the program.
\item {\em Architecture model.}
  First, we assume a ``linear'' execution model in each core of the processor.
  Informally, this means that the cost of executing a block of code followed by
  another is roughly equal to the sum of costs of executing the blocks
  individually.
  This assumption is not strictly true in most modern processors due to
  out-of-order and pipelined execution. 
  However, this effect is usually minor and localized when the two
  blocks of code are of sufficient size.
\item {\em Locking model.}
  We model the cost of acquiring a lock as a function of the contention
  for the lock, i.e., the number of threads waiting to acquire the lock.
  We denote by $\LockCost(k)$ the average cost of acquiring a lock when
  $k$ threads are waiting.
  Note that $\LockCost(k)$ does not include the time spent waiting for
  the lock to become available, but only the cost of the steps for
  acquiring the lock in the lock implementation, including cost of any
  system calls, inter-thread communication, etc.
  We extend the domain of $\LockCost(k)$ from $k \in \Naturals$ to $k
  \in \mathbb{R}_{\geq 0}$ by smoothly interpolating between the
  integers.
  For most lock implementations, $\LockCost(k)$ is low when only one
  thread is attempting to acquire a lock, i.e., $k \approx 1$
  and is much larger when $k >> 1$.
  
To measure $\LockCost(k)$, we run a concurrent program both
with and without locks under different contentions.  The difference between the
times gives us an estimate of the locking cost.
Note that these values need to be measured only once for each
computing platform. 
\end{compactitem}
%Before we proceed, we briefly describe how we measure $\LockCost(k)$ for
%different $k$. 
%Note that these values need to be measure only once for each
%architecture and implementation of locks.
%To measure $\LockCost(k)$, we run a concurrent program both with and
%without locks under different contentions.
%The difference between the times gives us a estimate of the locking
%cost.
%
\noindent{\bf\em Performance parameters.}
Given a fixed usage model and scheduler, our performance model predicts
the performance of a concurrent program based on several parameters of
the program.
Intuitively, fixing a usage model and a scheduler gives us a probability
distribution over the executions of the program, and program parameters
are then defined as the expected values of random variables over this
probability distribution.
Here, we define these parameters informally.
\begin{compactitem}
\item {\em Contention.} 
  We define the contention for a set of statements $\InstrSet$ as the
  average number of threads executing statements from $\InstrSet$ over
  time where we only consider time points when at least one thread is
  executing some statement in $\InstrSet$.
  We say that a thread is executing a lock statement even when it is
  waiting to acquire the lock.
  We usually denote contention using $\contention$.
\item {\em Cost.}
  The cost of a set of statements $\InstrSet$ is defined as the
  expected time spent per execution in executing statements from
  $\InstrSet$.
  We usually denote this parameter using lower case $\criticalCost$.
\item {\em Lock acquisitions.}
  The number of lock acquisitions is the expected number of {\tt lock}
  statements executed per execution by all the threads combined.
  We usually denote this parameter using $\LocksAcquired$.
\end{compactitem}

\noindent{\bf\em The performance model.}
First, note that the performance model is a statistical model and is not
formally guaranteed to always predict correctly which program performs
better. Like any statistical model, the suitability of the model is 
to be tested by experiment (this is done \secref{impl}).

The execution time of a program can roughly be split into three parts:
\begin{inparaenum}[(a)]
\item time taken to execute the lock-free code; 
\item time taken to acquire locks; and
\item time spent waiting to acquire locks.
\end{inparaenum}
In our model, (a)~is obtained (by profiling) as the cost $\criticalCost$ of the
relevant sections of code, and (b)~is obtained as the number of lock
acquisitions multiplied by the cost $\LocksAcquired$ of acquiring a lock.
The most important parameter that goes into predicting (c)~is
contention $\contention$ for the critical sections, i.e., how many other threads
are trying to execute the critical sections.
The most complex part of our model is the equation we use to predict
contention using the other parameters.
Our performance model is first constructed by measuring the above
parameters (contention, cost, and, locks acquisitions) for the solution
program $\cProg_C$ corresponding to the coarsest lock placement (see
Section~\ref{sec:syntactic_optimization}).

Let $\InstrSet$ be the set of all statements protected by any lock in
$\cProg_C$. 
We denote the contention for $\InstrSet$, the cost of $\InstrSet$, and
lock acquisitions in $\cProg_C$ by $\contention$, $\criticalCost$, and $\LocksAcquired$.
Now, consider a solution program $\cProg_R$ where the lock placement is a
refinement of the coarsest lock placement, i.e., every statement
is protected by a lock in $\cProg_R$ is also protected by a lock in
$\cProg_C$.
Let $\InstrSet'$ be the set of statements protected by any lock in
$\cProg_R$; we have that $\InstrSet' \subseteq \InstrSet$.
Let the cost of the $\InstrSet'$ and the lock acquisitions in $\cProg_R$ be
$\criticalCost'$ and $\LocksAcquired'$.
Our performance model predicts that the contention for the new critical
sections to be $\contention'$ where:
\[ 
  \contention' = 1 + (\contention-1) \cdot \frac{\mbox{Avg.\ time taken by one thread to
  execute $\InstrSet'$}}{\mbox{Avg.\ time taken by one thread to
    execute $\InstrSet$}}
\]
Intuitively, when a thread is executing $\InstrSet'$, on an average
there are $\contention - 1$ threads executing the coarser critical section as the
contention for $\InstrSet$ is $\contention$. 
For these $\contention-1$ threads executing $\InstrSet$, we approximate the
probability of each executing $\InstrSet'$ to be the fraction of time
each thread spends in $\InstrSet'$.
% Note that here, we assume that the contention for $\InstrSet$ does not
% change when using finer locks.

The average time taken by a thread to execute $\InstrSet$ in $\cProg_R$
can be written as the sum of the times taken to execute $\InstrSet'$ and
$\InstrSet \setminus \InstrSet'$.
Now, $\InstrSet \setminus \InstrSet'$ consists of only unprotected
statements which each thread can execute independently. 
Therefore, its cost is approximated as $(\criticalCost - \criticalCost')$.
We model the average time to execute $\InstrSet'$ as 
$(\criticalCost' + \LocksAcquired'*\LockCost(\contention')) * \max(\contention'-0.5, 1)$. 
Intuitively,
\begin{inparaenum}[(a)]
  \item $\criticalCost'$ is the cost of executing $\InstrSet'$ in the absence
    of other threads,
  \item $\LocksAcquired' * \LockCost(\contention')$ is the cost of acquiring the
    $\LocksAcquired'$ locks; and 
  \item the factor $\contention' - 0.5$ arises as each thread has to wait for $\contention'$
    others to finish executing the critical sections on an average.
    The $0.5$ correction term arises from a more careful analysis of
    what part of the critical section each thread is executing.
\end{inparaenum}
Hence, we get the following expression for contention:
\begin{equation}
  \label{eqn:contention}
  \contention' = 1 + (\contention-1) \cdot 
  \frac{(\criticalCost' + \LocksAcquired'*\LockCost(\contention')) * \max(\contention' - 0.5, 1)}
  {(\criticalCost - \criticalCost') + (\criticalCost' + \LocksAcquired'*\LockCost(\contention')) * \max(\contention' - 0.5, 1)}
\end{equation}
The rating given by the performance model to the program $\cProg_R$ is
defined to be the average time taken for a thread to execute.
Formally, we define
\begin{equation}
  \label{eqn:rating}
  \PerfModel(\cProg_R) = (\criticalCost - \criticalCost') + (\criticalCost' +
\LocksAcquired'*\LockCost(\contention')) * \max(\contention'-0.5, 1)
\end{equation}
As $\PerfModel(\cProg_R)$ is dependent only on $\criticalCost'$ and $\LocksAcquired'$, we abuse
notation by writing $\PerfModel(\criticalCost', \LocksAcquired')$ instead of $\PerfModel(\cProg_R)$.

\noindent{\bf\em Profiling model parameters.}
Here, we show how to profile $\cProg_C$ to obtain $\contention$, $\criticalCost$ and $\LocksAcquired$.
Further, we also measure some finer grained statistics so that $\criticalCost'$ and
$\LocksAcquired'$ can be constructed for all $\cProg_R$ with finer lock placements.

In practice, profiling code to measure execution times often changes the
time taken to execute the code. 
This effect is larger the smaller the size of the code being profiled.
Therefore, it is not practically possible to accurately measure the time
take to execute a single statement.
Hence, we divide the program into {\em blocks} of single-entry
multi-exit pieces of code.
While these blocks can be chosen arbitrarily, in practice, we
heuristically choose blocks such that they are likely to either be
completely contained in a critical section, or completely outside a
critical section.
This can be done by choosing the block boundaries to be in code which is
``uninteresting'' for concurrency.
For each statement $\instr$, let $\blk(\instr)$ represent the block it
belongs to.

Let $\cProg_C$ be the correct program with the coarsest possible locks obtained
from the optimization procedure in
Section~\ref{sec:syntactic_optimization} and let $\InstrSet$ be the set
of statements protected by a lock in $\cProg_C$.
We perform several kinds of profiling on $\cProg_C$ to obtain the following.
\begin{compactitem}
\item {\em Frequency measurement.} The number of times each statement
  $\instr$ is executed in each run is recorded, and these numbers are
  averaged over the runs to obtain $\freq(\instr)$.
  The frequencies will be used to obtain the average number of lock acquisitions
  for any given lock placement, i.e., the parameter $\LocksAcquired'$.
\item {\em Timing measurement.} For each block $\block$, the amount of
  time spent executing the block in each run is measured, and these
  times are averaged over the runs to obtain $\cost(\block)$.
  The cost of the $\InstrSet$ is approximated as $\sum_{\block \cap
  \InstrSet \neq \emptyset} \cost(\block)$.
  The block costs will be similarly used to obtain the average cost of
  the critical sections for any given lock placement.
\item {\em Contention measurement.} We measure the contention for
  $\InstrSet$ as follows: while the program is being executed, it is
  interrupted at various points and the number of threads executing
  $\InstrSet$ is recorded.
  Of these samples, the ones having no threads executing $\InstrSet$ are
  eliminated and the rest of the samples are averaged to get the
  contention $\contention$.
\end{compactitem}
Note that each of these measurements is performed on different profiling
runs as each measurement can interfere with the others.

Now, given $\cProg_R$, $\InstrSet$, and $\InstrSet'$ as in the previous
paragraph, we approximate $\criticalCost'$ and $\LocksAcquired'$ as follows:
\begin{compactitem}
\item The cost of $\InstrSet'$ is obtained by summing the cost of all
  blocks which are contained in $\InstrSet'$, i.e., $\criticalCost' = \sum_{\block
  \cap \InstrSet' \neq \emptyset} \cost(\block)$.
  If the blocks are chosen such that each block is either completely
  contained in $\InstrSet'$ or completely outside $\InstrSet'$, this
  approximation of $\criticalCost'$ is precise.
  Note that obtaining $\criticalCost'$ by simply summing up the block costs uses the
  assumption that the execution model is linear.
\item The average number of lock acquisitions $\LocksAcquired'$ is obtained by
  summing the frequencies of all statements which immediately follow a
  {\tt lock} statement, i.e., $\LocksAcquired' = \sum_{\instr} \freq(\instr)$, 
  where $\instr$ ranges over statements following {\tt lock} statements.
\end{compactitem}

\noindent{\bf\em Obtaining the optimal program.}
First, we describe how to extend the global lock placement constraint
$\smt$ with additional variables $\criticalCost'$ and $\LocksAcquired'$ that represent the cost
of critical sections and average number of lock acquisitions.
Further, we introduce auxiliary variables $b_i$ to represent
whether $\block_i \cap \InstrSet' \neq \emptyset$ and variables
$t_{\instr}$ to represent if $\instr$ is the first statement following a
lock statement. 
Now, we can constrain $\criticalCost'$ as follows:
\begin{gather*}
  \bigwedge_{\instr} \InLock(\instr) \implies \blk(\instr) \\
  \criticalCost' = \sum_i (\ITE{b_i}{\langle \cost(\block_i) \rangle}{0}) \\
\end{gather*}
Similarly, for average lock acquisitions $\LocksAcquired'$, we have:
\begin{gather*}
  \bigwedge_{\instr} \bigvee_{lk} \LockBefore(\instr, lk) \implies t_{\instr} \\
  \LocksAcquired' = \sum_{\instr} (\ITE{t_{\instr}}{\langle \freq(\instr) \rangle}{0}) \\
\end{gather*}
In the above constraints, $\langle \cost(\block_i) \rangle$ and $\langle
\freq(\instr) \rangle$ are not symbolic expressions, but the concrete values
measured by profiling.
We denote the formula obtained by taking the conjunction of $\smt$ and
the above constraints as $\widehat{\smt}$.

We now present a procedure to generate the optimal lock placement
according to the performance model.
Intuitively, we solve Equations~\ref{eqn:contention}
and~\ref{eqn:rating} numerically to obtain the $\PerfModel$ rating for 
for the full range of values for $\criticalCost'$ and $\LocksAcquired'$ to build a
map from $\criticalCost'$ and $\LocksAcquired'$ values to $\PerfModel$
values.
Then, using an SMT solver, we search for lock placements for which the
$\criticalCost'$ and $\LocksAcquired'$ values fall in the neighbourhood of minima in this
map.

Formally, we define a {\em region} $\Region$ to be an expression of the
form $\criticalCost_1 \leq \criticalCost' \leq \criticalCost_2 \wedge \LocksAcquired_1 \leq \LocksAcquired' \leq \LocksAcquired_2$ where $\criticalCost_1, \criticalCost_2,
\LocksAcquired_1, \LocksAcquired_2 \in \mathbb{R}$.
We define $w(\Region)$ and $h(\Region)$ to be $\criticalCost_2 - \criticalCost_1$ and
$h(\Region) = \LocksAcquired_2 - \LocksAcquired_1$, respectively.
For each rectangle, we define the query $\UniquePerformance(\Region)$ to
return true if and only if all models of $\smt \wedge \Region$ have the
same values for $\criticalCost'$ and $\LocksAcquired'$.

Algorithm~\ref{algo:perf_opt} presents a procedure to find the 
model of $\widehat{\smt}$ corresponding to the lock placement with the
minimal $\PerfModel$ value.
It takes the following additional parameters:
\begin{inparaenum}[(a)]
\item $\Delta_1$ and $\Delta_2$ which bound the rate of change of
  $\PerfModel(\criticalCost', \LocksAcquired')$ with respect to $\criticalCost'$ and $\LocksAcquired'$, and
\item a shattering parameter $k$.
\end{inparaenum}
Any integer $k > 1$ can be passed as the shattering parameter without
affecting the correctness of the algorithm--in practice, we use $k =
10$.
In practice, the bounds $\Delta_1$ and $\Delta_2$ can be approximated
numerically.

The algorithm proceeds by building a set of {\em concrete points} placed
at regular intervals across the full space of values of $\criticalCost'$ and $\LocksAcquired'$,
and computing $\PerfModel(\criticalCost', \LocksAcquired')$ for these points.
Note that these concrete points do not necessarily correspond to models
of $\widehat{\smt}$, but they will be used to guide the search for
optimal models.
We ask the SMT solver to generate models in the region containing the
best concrete point.

The full space (and the initial region) is given by $(0 \leq
\criticalCost' \leq \criticalCost) \wedge (0 \leq \LocksAcquired' \leq
\LocksAcquired_{max})$ where $\LocksAcquired_{max}$ is $\sum_{\instr}
\freq(\instr)$.
In each step, the region $\Region$ into which the concrete point with
the minimum $\PerfModel$ value belongs to is picked.
If $\Region$ does not contain a model, i.e., no locking
scheme has $\criticalCost'$ and $\LocksAcquired'$ in $\Region$, it is deleted; further, all
concrete points falling into $\Region$ are also deleted.
If the region does contain a model and the model's $\PerfModel$ value is 
less than the minimum, we record the minimum and the model.
Further, we eliminate all other regions where all points have
performance values larger than the current minimum.
To do this, the algorithm first picks the concrete point with the lowest
$\PerfModel$ values in each region and subtracts from this value the
height and width of the region multiplied by parameters $\Delta_1$ and
$\Delta_2$.
If this value is greater than $\MinPerf$, the region is removed.
Note that $\Delta_1 \cdot w(\Region') + \Delta_2 \cdot h(\Region')$ is
maximum possible variation of the $\PerfModel$ function within the
region $\Region'$ as $\Delta_1$ and $\Delta_2$ are upper bounds for 
$\vert \frac{\partial \PerfModel}{\partial \criticalCost'}  \vert$ and $\vert
\frac{\partial \PerfModel}{\partial \LocksAcquired'}  \vert$.

As the last part of the step, if the region contains more models with
distinct $\criticalCost'$ and $\LocksAcquired'$, we {\em shatter} the region, i.e., replace the
region with $k^2$ smaller sub-rectangles of equal size.
For these new regions, for every region that does not contain a concrete
point, we pick a concrete point from the region and add it to $Pts$. 

\begin{algorithm}
  \caption{Searching of the optimal program}
  \label{algo:perf_opt}
  \begin{algorithmic}
    \Require Global lock placement constraint $\widehat{\smt}$
    \Require Performance model parameters $\criticalCost$, $\LocksAcquired_{max}$
    \Require Shattering parameter $k > 1 \in \Naturals$
    \Require Upper bounds $\Delta_1$ and $\Delta_2$ on the partial
    derivatives $\vert \frac{\partial \PerfModel}{\partial \criticalCost'}
    \vert$ and $\vert \frac{\partial \PerfModel}{\partial \LocksAcquired'} \vert$
    \Ensure Minimal model $\MinModel$ of $\widehat{\smt}$ 
    \State $Pts \gets \{ (\criticalCost', \LocksAcquired', \PerfModel(\criticalCost', \LocksAcquired')) | (0 \leq \criticalCost' = \frac{\criticalCost \cdot k_1}{2k} \leq \criticalCost)~\wedge  $ 
      \flushright $ (0 \leq \LocksAcquired' = \frac{\LocksAcquired_{max} \cdot k_2}{2k} \leq
      \LocksAcquired_{max}) ~\mbox{ for $k_1, k_2 \in \Naturals$ }\}$ 
    \State \flushleft $InitRegion \gets (0 \leq \criticalCost' \leq \criticalCost) \wedge (0 \leq \LocksAcquired' \leq \LocksAcquired_{max})$
    \State $Regions \gets \{ InitRegion \}$
    \State $\MinPerf \gets \infty  ; \MinModel \gets None$
    \While { $Regions \neq \emptyset$ }
      \State $\Region \gets GetRegion(GetMinimum(Pts))$
      \State $Regions \gets Regions \setminus \{ \Region \}$
      \If { $\widehat{\smt} \wedge \Region$ is not satisfiable }
        \State $Pts \gets Pts \setminus \{ Pt | Pt \in \Region \}$
        \State {\textbf{continue}}
      \EndIf
      \State $M \gets GetModel(\widehat{\smt} \wedge \Region)$
      \If { $\MinPerf >  \PerfModel(M[\criticalCost'], M[\LocksAcquired'])$}
        \State $\MinPerf \gets \PerfModel(M[\criticalCost'], M[\LocksAcquired'])$
        \State $\MinModel \gets M$
      \EndIf
      \If { $\neg \UniquePerformance(\Region)$ }
        \State $\Regions \gets \Regions \cup Shatter(\Region) $
        \For {$\Region' \in \Regions$}
          \If {$\not \exists (\criticalCost', \LocksAcquired', p) \in Pts : (\criticalCost', \LocksAcquired')\in \Region'$}
            \State {Pick $(\criticalCost', \LocksAcquired') \in \Region$}
            \State $Pts \gets Pts \cup \{ (\criticalCost', \LocksAcquired', \PerfModel(\criticalCost', \LocksAcquired')) \}$
          \EndIf
        \EndFor
      \EndIf
      \For { $\Region' \in \Regions$ }
        \If { $\forall (\criticalCost', \LocksAcquired', p) \in Pts : (\criticalCost', \LocksAcquired') \in \Region'$
          $\wedge (p - \Delta_1 \cdot w(\Region') - \Delta_2 \cdot h(\Region')) > \MinPerf $ }
          \State $Regions \gets Regions \setminus \{ \Region' \}$
          \State $Pts \gets Pts \setminus \{ Pt | Pt \in \Region' \}$
        \EndIf
      \EndFor
    \EndWhile
    \Return $\MinModel$
  \end{algorithmic}
\end{algorithm}

\begin{theorem}
  If $\widehat{\smt}$ is satisfiable, Algorithm~\ref{algo:perf_opt}
  returns the model corresponding to the lock placement that yields the
  optimal program as per the objective function $\PerfModel$.
\end{theorem}

\noindent{\bf\em Extension to multiple locks.}
The performance model presented above is for the case where only lock
variable is used.
However, it can be easily extended to multiple locks by defining
separate contention variables for different critical section protected
by different locks.
Equation~\ref{eqn:contention} is replaced by multiple equations for
contention for each lock $lk$ where, again, the denominator
corresponds to the average time a thread spends executing the coarse
grained critical section, and the numerator corresponds to the average
time a thread spends executing the critical section protected by
$lk$.

\comment{
\noindent{\bf\em Hybrid optimization criteria.}
There is an interesting class of hybrid optimization criteria that can
be obtained by combining the frequency measurement obtained by profiling
with syntactic criteria from Section~\ref{sec:syntactic_criteria}.
For example, consider the coarsest grained locks 
\arsays{Something something something, but we do not speak of such
things}
}

\section{Implementation and Evaluation}
\label{sec:impl}

\noindent{\bf\em Implementation}

\lrsays{TODO: discuss deadlocks}
\rs{Also, Anonymize}
We implemented the optimal lock placement technique, described 
above, in a tool called \ourtool.  \ourtool\ is based on the open 
source \pasttool\ synthesis tool~\cite{CAV15}.  \pasttool\ uses 
language-inclusion-based conflict detection \rs{we don't mention 
race detection
anywhere else. So I would skip.}\lrsays{is this better?} in 
conjunction with data-oblivious abstraction, and implements a 
greedy algorithm for lock placement.  We replaced the greedy 
algorithm with the optimal lock placement algorithm.  In addition, 
\pasttool\ does not guarantee deadlock freedom of synthesized 
solutions, whereas our constraint-based synthesis algorithm does.

\ourtool\ is comprised of 5400 lines of C++ code, including 775 
lines that implement the optimal lock placement procedure.  It 
uses Z3 version 4.4.1 (unstable branch) as MaxSMT solver.  We use 
Clang/LLVM 3.6 for parsing and generating C code.  In particular, 
we rely on the Clang's rewriter to insert text into the original 
source file. This has the advantage that the output file preserves 
original source file content and formatting.

%It is available as open-source software along with the benchmarks at XXX.

\noindent{\bf\em Benchmarks}

We evaluate \ourtool\ using two benchmarks.  The first benchmark, 
proposed by the \pasttool\ project, is based on the Linux driver 
for the TI CPMAC Ethernet controller device~\cite{CAV15}.  To the 
best of our knowledge, this benchmark represents the most complex 
synchronization synthesis case study, based on real-world code, 
reported in the literature.  The CPMAC benchmark is a \emph{bug 
fixing} benchmark---most synchronization code is already in place 
in the driver; however preemption safety may be violated due to 
missing locks.  The goal of synthesis is to detect and 
automatically correct such defects.  

In our experiments we focus on the quality of synthesized code: 
out of multiple correct solutions we would like to pick one that 
is consistent with the user-preferred locking strategy.  This can 
be formalized as an optimal synthesis problem.  The two locking 
strategies used in device drivers in practice are 
\emph{fine-grained} and \emph{coarse-grained} locking.  The former 
requires placing as few statements as possible inside locks and 
using different locks to protect independent operations.  This 
strategy is used in conjunction with low-overhead spin locks and 
is particularly useful for protecting data accessed by interrupt 
handler functions, as waiting inside an interrupt handler can 
severely degrade the overall system performance.  The coarse 
locking strategy is used to protect non-performance-critical 
operations, where code clarity is more important than 
parallelizability.
\ttsays{Could the coarsest not also perform better due to less lock statements?}
These strategies correspond to the $\Coarsest$ 
and $\Finest$ optimization criteria defined in 
Section~\ref{sec:syntactic_optimization}.

%We formalize the fine-grained discipline using a cost function 
%that adds a penalty of one for every program location that is 
%protected by a synthesized lock.  We refer to it as the 
%\emph{minimal lock scope} objective (\emph{minscope} for short), 
%as it places as few statements as possible inside critical 
%sections.  To further ensure that separate locks are synthesized 
%to address independent program races, we add a high penalty ($100$ 
%in our experiments) for every pair of conflicts that involve 
%disjoint code sections but are eliminated using the same lock 
%variable.  We call the resulting objective the \emph{fine-grained 
%locking} objective, or \emph{fine}, as it captures the intuitive 
%meaning of the fine-grained lock placement.  
%
%Finally, we enforce coarse-grained locking by adding to the 
%minscope objective a high penalty for every lock statement placed 
%in the code.  The resulting objective function, called 
%\emph{coarse}, merges several critical sections when possible, 
%while still avoiding locking statements unnecessarily.

Our second benchmark explores a very different scenario where we 
would like to synthesize locking for a complete program module 
rather than fix defects in existing lock placement.  Furthermore, 
optimal locking strategy in this benchmark depends on the hardware 
platform and workload and can only be obtained using the dynamic 
cost model described in Section~\ref{sec:profiling_optimization}.  
We envisage a usage model where the developer ships a version of 
software for a non-preemptive scheduler, and lock placement is 
synthesized as part of the software integration process with the 
help of profiling data collected for a particular target platform 
and workload.

Specifically, we consider the \emph{memcached} in-memory key-value 
store server version 1.4.5~\cite{memcached}.  The core of 
memcached is a non-reentrant library of store manipulation 
primitives.  This library is wrapped into the \src{thread.c} 
module that implements a thread-safe API used by server threads.  
Each API function performs a sequence of library calls protected 
with locks. 

Table~\ref{t:benchmarks} summarizes our benchmarks in terms of the 
number of lines of code and the number of concurrent threads in 
each benchmark.  We consider six variants of the CPMAC driver, 
five containing different concurrency-related defects and one 
without defects.

\begin{table*}[t]
\footnotesize
\newcommand{\notime}{\textless 1s}
\renewcommand{\thefootnote}{{\it\alph{footnote}}}
\small
\begin{tabular}{|l || >{\raggedright}p{0.07\textwidth} | p{0.07\textwidth} | p{0.07\textwidth} | p{0.07\textwidth} | p{0.07\textwidth} | p{0.07\textwidth} | p{0.07\textwidth} | p{0.07\textwidth} | p{0.09\textwidth} |}
\hline
\multirow{3}{*}{Name}         & \multirow{3}{*}{LOC} & \multirow{3}{*}{Threads} & \multicolumn{5}{c|}{Time (s)}                                                         & \multirow{3}{0.09\textwidth}{Memory (MB)} \\
\cline{4-8}
                              &                      &                          & Conflict          & \multicolumn{4}{c|}{Objective function}                           &                              \\
\cline{5-8}
                              &                      &                          & analysis          & None         & $\Coarsest$ & $\Finest$ & model-based &                              \\
\hline\hline
cpmac.bug1                    & 1275                 & 5                        & 6.0               & 0.21         & 1.59        &  1.13     &  -          & 156                          \\
cpmac.bug2                    & 1275                 & 5                        & 150.4             & 0.80         & 63.0        &  41.37     &  -          & 1210                         \\
cpmac.bug3                    & 1270                 & 5                        & 11.2              & 0.56         & 16.18       &  9.64    &  -          & 521                          \\
cpmac.bug4                    & 1276                 & 5                        & 107.4             & 0.59         & 10.54       &  6.50    &  -          & 5392                         \\
cpmac.bug5                    & 1275                 & 5                        & 137.3             & 0.57         & 11.03       &  7.67    &  -          & 3549                         \\
cpmac.correct\footnotemark[1] & 1276                 & 5                        & 2.1               & -            & -           & -        &  -          & 114                          \\
memcached                     & 294                  & 2                        & 49.51             & 0.685        & 6.205       &  2.10     &  23.0       & 114                          \\
\hline

\end{tabular}
%\footnotemark[1] initial bound~~~~
\begin{flushleft}
\footnotemark[1] bug-free example (\ourtool\ stops after checking preemption safety) \\
\end{flushleft}
\caption{Benchmarks.}
\label{t:benchmarks}
\end{table*}

\noindent{\bf\em Evaluation}

Our experiments aim to evaluate three main properties of the 
optimal synthesis procedure: (1) \emph{efficiency} of the 
synthesis algorithm, (2) \emph{correctness} and (3) \emph{quality} 
of synthesized lock placement.  Table~\ref{t:benchmarks} reports 
on the efficiency of the algorithm by showing the time spent (1) 
analyzing the input program and computing the set of conflicts, 
(2) synthesizing a lock placement without an objective function, 
i.e., by solving the set of hard constraints only, and (3) 
computing optimal lock placement for coarse-grained and fine-grained
objective functions and, for the memcached benchmark only, 
the objective function based on performance model.  It 
also shows the peak memory consumption of \ourtool, which is in 
all cases reached during anti-chain-based language inclusion 
check.  All measurements were done on an Intel core i5-3320M 
laptop with 8GB of RAM. In all experiments, synthesis completed 
within a few minutes or less.  Importantly, the added cost of 
optimal lock placement is modest compared to the analysis step. 

Next, we evaluate correctness of synthesized lock placement.  In 
all examples, \ourtool\ enforced both preemption equivalence and 
deadlock-freedom of the synthesized solution.  In contrast, 
previous tools, including \pasttool, may introduce deadlocks 
during synthesis.  This is not just a theoretical possibility: 
in one of the examples (\src{cpmac.bug2}), \pasttool\ 
synthesizes a solution that contains a deadlock.  While \pasttool\ is
able to detect such deadlocks, it leaves it to the user to fix them.

Next, we focus on the quality of synthesized solutions.  
Table~\ref{t:locks} compares the implementation synthesized using 
each objective functions in terms of (1) the number 
of locks used in synthesized code, (2) the number of lock and 
unlock statements generated, and (3) total number of program 
statements protected by synthesized locks.

\begin{table*}[t]
\footnotesize
\renewcommand{\thefootnote}{{\it\alph{footnote}}}
\small
\begin{tabular}{|l || p{0.05\textwidth}  p{0.05\textwidth}  p{0.05\textwidth} | p{0.05\textwidth} p{0.05\textwidth} p{0.05\textwidth} | p{0.05\textwidth} p{0.05\textwidth} p{0.05\textwidth} |}
\hline
\multirow{3}{*}{Name}         & \multicolumn{3}{c|}{No objective} & \multicolumn{3}{c|}{$\Coarsest$} & \multicolumn{3}{c|}{$\Finest$} \\
\hline
                              & locks                & locks/ unlocks        & protected instr   & locks                & locks/ unlocks        & protected instr   & locks                & locks/ unlocks    & protected instr   \\
\hline\hline
cpmac.bug1                    & 2 & 6/6   & 11 & 1 & 3/3 & 11 & 1 & 3/3 & 9  \\
cpmac.bug2                    & 2 & 22/23 & 119& 1 & 4/4 & 98 & 1 & 6/7 & 95 \\
cpmac.bug3                    & 1 & 4/4   & 29 & 1 & 2/3 & 29 & 1 & 5/6 & 28 \\
cpmac.bug4                    & 4 & 16/16 & 53 & 1 & 4/4 & 53 & 1 & 6/6 & 26 \\
cpmac.bug5                    & 3 & 15/15 & 30 & 1 & 4/4 & 30 & 1 & 5/5 & 30 \\
memcached                     & 2 & 5/5   & 26 & 1 & 1/1 & 28 & 1 & 2/2 & 24 \\
\hline

\end{tabular}
%\footnotemark[1] initial bound~~~~
\begin{flushleft}
\end{flushleft}
\caption{Lock placement statistics: the number of synthesized lock variables, lock and unlock statements, and the number of abstract statements protected by locks for different objective functions.}
\label{t:locks}
\end{table*}

Interestingly, even though in CPMAC benchmarks we start with a 
program that has most synchronization already in place, different 
objective functions produce significantly different results in 
terms of the size of synthesized critical sections and the number 
of lock and unlock operations guarding them: the fine-grained
objective synthesizes smaller critical sections at the cost of
introducing a larger number of lock and unlock operations.  The
implementation synthesized without an objective function is clearly
of lower quality than the optimized versions: it contains large 
critical sections, protected by unnecessarily many locks. 

%The fine-grained objective introduces multiple 
%locks to minimize false contention among different critical 
%sections, but, as a result, generates more lock statements, as 
%some of the critical sections must be protected by multiple locks.

Finally, we evaluate the profiling-based synthesis method on the 
memcached benchmark.  The key question here is whether the 
performance model built using profiling data accurately predicts 
program performance for various lock placements and, hence, leads 
to a synthesized implementation with optimal performance.

Preliminary analysis showed that the high-level structure of the 
benchmark resembles the program in Figure~\ref{ex:worksharing}: 
each memcached operation consists of several phases that access 
the shared data store and must be locked to enforce atomicity.  In 
between atomic sections, the lock can be safely dropped; however 
in most cases this does not make sense, as the lock must be 
re-acquired after a few statements.  The only exception is the 
store update operation where two atomic sections are separated by 
a memory copy operation between two thread-local variables, where 
the size of data copied is equal to the size of the new data item 
written to the store.  If the average item size is large enough, 
then the performance gain due to unlocking the copy operation and 
running it concurrently with other threads outweighs the overhead 
of the extra lock operations.  

Thus, the benchmark effectively allows two valid lock placements: 
the coarse grained placement, which in this case locks the entire 
body of the server loop, and a finer-grained version that drops 
the lock during the copy operation.  The choice of the optimal 
placement depends on the average data item size.  In a real-world 
scenario, this depends on the kind of data stored in the given 
memcached installation, e.g., microblog entries, email messages, 
or high-resolution images. 

In order to construct a performance model for the memcached 
benchmark, we developed an artificial workload that generates a 
randomized sequence or read, write, and update requests to 
memcached, parameterized by the average item size.  We pick $12$ 
different parameter values, ranging from $32$ to $65,536$ bytes 
and for each of them measure model parameters following the 
methodology outlined in Section~\ref{sec:profiling_optimization}: 
we break the program into blocks and measure $\LockCost$, $\contention$, 
$\criticalCost$, $\LocksAcquired$, and, for each block, $\cost(\block)$.  All measurements were done on a
quad-core Intel Core i7-3520M machine, running one server thread 
per core, which is the standard memcached configuration.  All 
steps of the synthesis process, including profiling, MaxSMT 
constraint generation and lock placement in the source code are 
automated, with the exception of program decomposition into blocks 
(Section~\ref{sec:profiling_optimization}), which is currently 
done manually, but can be readily automated.

\begin{figure}[t]
\includegraphics[trim = 0.0in 0.0in 0.0in 0.0in,clip,width=0.75\linewidth]{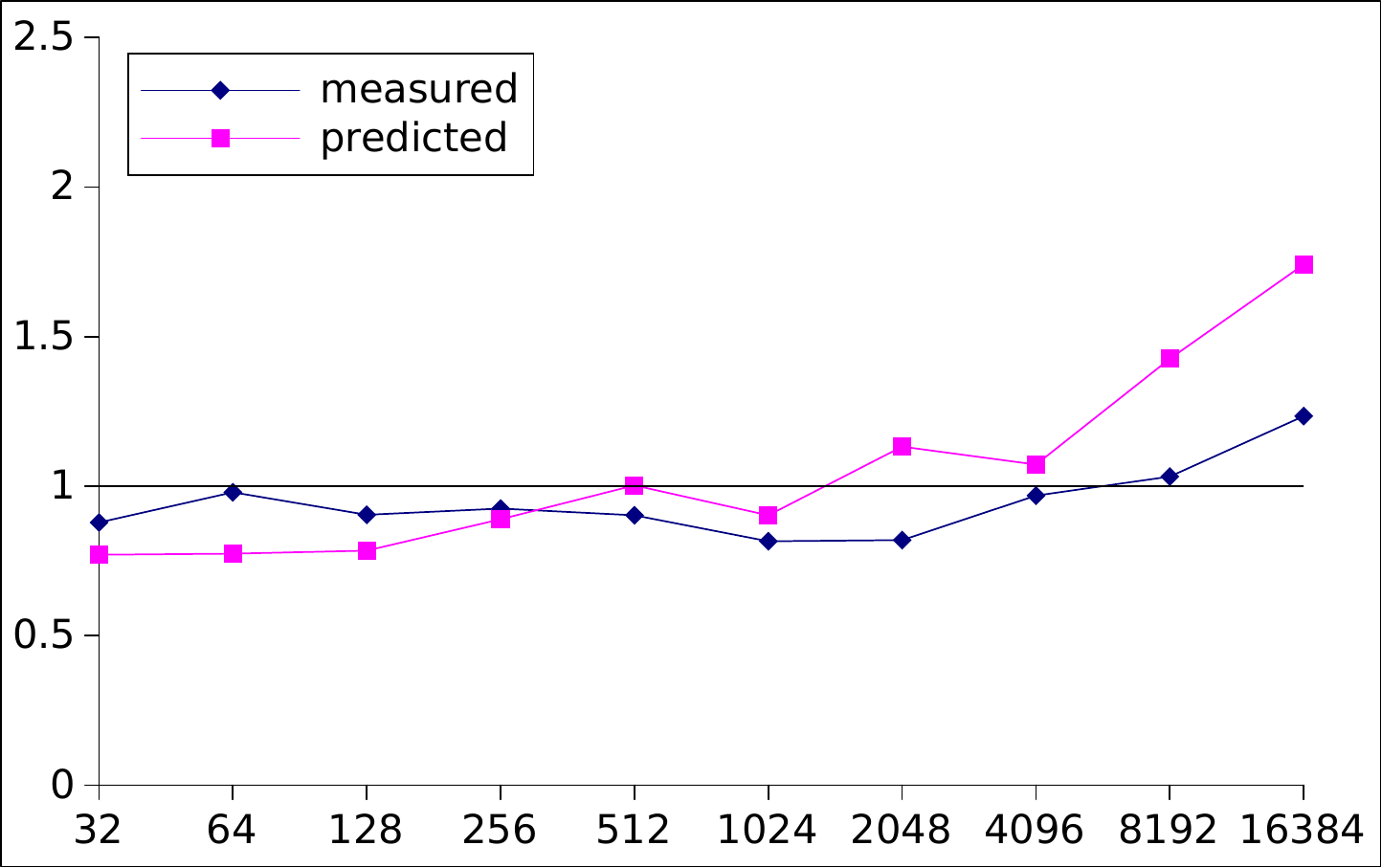}  
\caption{Precision of the statistical model.}
\label{fig:memcached}
\end{figure}

We evaluate our statistical model by extracting predicted 
performance of the finer-grained lock placement for different 
parameter values and comparing it against actual measured 
performance.  More precisely, we measure predicted and actual 
speed-up due to finer-grained locking relative to coarse-grained 
locking: $\PerfModel(\cProg_C)/\PerfModel(\cProg_R)$.  We plot the 
results in Figure~\ref{fig:memcached}.  
Values less than $1$ mean 
that coarse locking performs better than finer locking; 
conversely, values $>10$ .  

In all cases, predicted performance is close to the actual 
performance.  In $8$ out of $11$ data points, both lines are 
either above or below $1$, and the lock placement synthesized by 
\ourtool\ for these parameter values is indeed optimal.  

%For the remaining values our predicted classification is wrong, 
%leading to suboptimal lock placement.  However, the performance of 
%synthesized code is only marginally below optimal in this case.

%% In all cases, predicted performance is within XXX\% of the actual 
%% performance.  In XXX out of XXX data points, both lines are either 
%% above or below $1$, and the lock placement synthesized by 
%% \ourtool\ for these parameter values is indeed optimal.  For 
%% borderline values XXX our predicted classification is wrong, 
%% leading to suboptimal lock placement.  However, the performance of 
%% synthesized code is only marginally below optimal in this case.
%
These encouraging results experimentally justify our statistical 
performance model and the overall synthesis approach based on a 
combination of static program analysis and runtime profiling.

\bibliographystyle{plain}
%\bibliography{references}
\bibliography{refs}

\end{document}